\documentclass[aip,jmp,floats,amssymb,reprint]{revtex4-1}
\usepackage{bm}
\usepackage{amsmath,amsfonts,amsthm}
\usepackage{mathtools}
\usepackage{hyperref}
\usepackage{color}

\newcommand{\beq}{\begin{equation}}
\newcommand{\eeq}{\end{equation}}
\newcommand{\beqa}{\begin{eqnarray}}
\newcommand{\eeqa}{\end{eqnarray}}
\newcommand{\beqan}{\begin{eqnarray*}}
\newcommand{\eenan}{\end{eqnarray*}}

\newcommand{\Dslash}{{\slash{\kern -0.5em}\partial}}
\newcommand{\Aslash}{{\slash{\kern -0.5em}A}}

\def\sqr#1#2{{\vcenter{\hrule height.#2pt
     \hbox{\vrule width.#2pt height#1pt \kern#1pt
        \vrule width.#2pt}
     \hrule height.#2pt}}}

\def\thinspace{\kern .16667em}

\def\xp{x_{{\kern -.2em}_\perp}}
\def\subp{_{{\kern -.2em}_\perp}}

\def\defeq{:{\kern -0.5em}=}
\def\Tr{{\rm Tr}\,}

\newcommand\Hilb{{\mathfrak H}}
\newcommand\States{{\mathfrak S}}
\newcommand\X{{\mathcal X}}
\newcommand\Dns{\mathrm{Dens}}
\newcommand\Potl{\mathrm{Potl}\,}
\newcommand{\Rc}{{\mathcal R}}
\newcommand{\Vc}{{\mathcal V}}
\newcommand{\Vcb}{\overline{\mathcal V}}
\newcommand{\PED}{{\mathcal{PE}}}

\def\E{{\mathcal E}}

\newcommand\Part{{\mathfrak P}}
\newcommand\Star{{{}^\star}}
\def\Std#1{ {}^\circ{#1}}

\newtheorem{lem}{Lemma}[section]

\newtheorem{cor}{Corollary}[section]

\newtheorem*{cor*}{Corollary}
\newtheorem*{thm*}{Theorem}
\newtheorem{prop}{Proposition}[section]

\begin{document}

\title{Coarse-grained spin density-functional theory:
infinite-volume limit via the hyperfinite}
\author{Paul~E.~Lammert}
\affiliation{
Dept. of Physics, 104B Davey Lab \\
Pennsylvania State University \\
University Park, PA 16802-6300}
\date{\today}

\begin{abstract}
Coarse-grained spin density functional theory (SDFT) is a 
version of SDFT which works with number/spin densities 
specified to a limited resolution --- averages over cells 
of a regular spatial partition --- and external potentials 
constant on the cells. This coarse-grained setting facilitates
a rigorous investigation of the mathematical foundations which
goes well beyond what is currently possible in the conventional
formualation.
Problems of existence, uniqueness and regularity of
representing potentials in the coarse-grained SDFT setting
are here studied using techniques of (Robinsonian) nonstandard analysis. 
Every density which is nowhere spin-saturated is
V-representable, and the set of representing potentials
is the functional derivative, in an appropriate generalized sense,
of the Lieb internal energy functional.
Quasi-continuity and closure properties of the set-valued
representing potentials map are also established.
The extent of possible non-uniqueness is similar to that 
found in non-rigorous studies of the conventional theory, 
namely non-uniqueness can occur for
states of collinear magnetization which are eigenstates of
$S_z$.
\end{abstract}
\maketitle

\section{Introduction}
\label{intro}

Modern electronic density functional 
theory\cite{Parr+Yang,Dreizler+Gross,Eschrig,Koch+Holthausen,vanLeeuwen03,LNP620,Engel+Dreizler,Capelle06}
(DFT) is a very successful basis for the computation of
many ground state properties used by chemists, physicists and materials 
scientists. However, it is not just practical computational algorithms, 
but also a distinctive way of looking at quantum many-body systems with its
own set of fascinating basic questions.
 
While the computational side of DFT is highly developed, the mathematical
foundations are relatively impoverished.
The fundamental questions revolve around the concept of 
$V$-representability. For a system of $N$ electrons,
with their mutual Coulomb repulsion, a single-particle density
$\rho({x})$ is said to be pure-state V-representable if 
some external single-particle potential $v({x})$ has a ground
state wavefunction with single-particle density $\rho$. It is mixed-state
V-representable (simply ``V-representable'' here)
if there is a mixed ground state (density matrix)
with density $\rho$. 
Natural questions are those of existence, uniqueness
and regularity of the representing potentials as a function of
density. The realization that some densities are not
pure-state V-representable, but are mixed-state V-representable
was responsible for the rise to prominence of the latter 
concept\cite{Levy79,Levy82,Lieb83} around 1980. About the same time,
some densities which are not even mixed-state V-representable
were suggested\cite{Englisch+Englisch-83}. 
Although V-representable densities are dense in an appropriate
topology, so are non-V-representable densities and there is as
yet no nontrivial positive characterization of V-representability.
Concerning uniqueness, the original paper of Hohenberg and 
Kohn\cite{Hohenberg+Kohn} already contained a strong uniqueness
result (modulo a trivial constant shift of the potential). 
Subsequently, when the theory was generalized to spin density functional
theory (SDFT) in order to study magnetic phenomena,
it was realized\cite{VonBarth+Hedin} that there was some breakdown 
of uniqueness in that context. There has recently been 
clarification\cite{VonBarth+Hedin,Capelle+Vignale,Eschrig+Pickett,Ullrich05,Gidopoulos07}
of the extent of that nonuniqueness.
Questions of regularity --- how much will the representing potential
of a density resemble those of nearby densities --- has been almost
entirely neglected. And not just by a lack of results; even the question
seems largely unacknowledged.
These foundational issues are also relevant to the computational 
side of DFT. Computability is threatened not only by lack of existence,
but also by lack of regularity. 

Chayes, Chayes and Ruskai\cite{CCR85} studied DFT for a lattice version of 
quantum mechanics and showed that, although serious mathematical difficulties or 
even pathologies may 
arise from the presence of arbitrarily short distance scales, the infinite-volume 
limit is a tractable problem. 
In particular, it was shown that {\em every} density is
a ground-state density of an essentially unique potential.
On a lattice, short distance scales are {\em completely} eliminated;
this is a fundamentally different quantum mechanics than the orthodox
continuum version.
Another way to keep short-distance-scale degrees of freedom
from causing problems without altering the underlying continuum quantum
mechanics was introduced as coarse-grained DFT\cite{Lammert06}.
The idea is that we only allow ourselves to specify densities
with some limited spatial resolution.  On scales finer than the resolution, 
the density is automatically relaxed by an energetic criterion. 
This approach seems much in the spirit of DFT. Conventional density functional
theory asks for the lowest-intrinsic-energy (kinetic plus Coulomb) state
consistent with a fully and exactly specified density. 
But it is entirely natural, and perhaps more computationally relevant,
to consider incomplete specifications.
Unique V-representability  holds also in this formulation and further 
regularity results were recently\cite{Lammert10a,Lammert10b} demonstrated. 
Although the infinite-volume limit was handled, it was not found to be
trivial, and for each property had to be approached anew.
Since practical computations work with limited representational resources,
usually in a form that amounts to limited spatial resolution,
the coarse-grained formulation might be considered a more suitable
grounding than the conventional fine-grained one. 
On the other hand, the coarse-graining scale can be taken as small
as one wishes ($10^{-30}$ m, say), so the approach is not
inherently imprecise.
Still, the infinite-resolution limit in which the coarse-graining scale 
goes to zero is of interest to understand the scale-dependences and
to make contact with things formulated in a conventional
infinite-resolution form such as an exact Coulomb potential.
For that limit the coarse-grained formulation seems to have an
advantage over the lattice, since a finer scale on the lattice 
requires a new lattice, not just some extra resolution. 
Some progress in that vein has been made\cite{Lammert10a,Lammert10b}
for ordinary (non-spin) DFT.  This paper has no new contribution 
concerning that problem, but instead returns to the single-scale 
setting and the infinite-volume limit for spin-density functional theory.

In previous work on the coarse-grained approach, it was observed that for a
system confined to a finite box the situation really is simple.
This paper aims to easily obtain the infinite-volume limit by
moving from finite boxes to the intermediate stage of a {\it hyperfinite} box.
``Hyperfinite'' is meant here in the sense of nonstandard analysis 
(NSA)\cite{Robinson66,Lindstrom85,Goldblatt,Loeb+Hurd,Loeb,Davis,AFHL,Stroyan+Luxemburg}
which is a rigorous way to use the notion of {\it infinitesimals} as well
as their reciprocals.
In that setting, a hyperfinite box can be asserted meaningfully to be
larger than the infinite-volume limit, yet formal properties of a
system confined to such a box can be deduced immediately from those
in a finite box. The added clarity brought by the infinitesimal methods
allows not just ordinary DFT, but also spin density functional theory
to be handled. Everything is done from scratch; there is no dependence
on the earlier results.  Thus, this paper also aims to promote the use of
infinitesimal methods, which are not nearly as well-known as 
they deserve.  It is hoped that nonstandard analysis will in the future 
allow the continuum limit also to be dealt with.

The next section reviews some basic ideas of SDFT and
\S \ref{basic-NSA} is a very brief introduction to nonstandard 
analysis (NSA). Readers with appropriate background can
skip these sections.
\S \ref{lsc+ns} gets down to work, proving some fundamental
lower semicontinuity and nearstandardness results using 
infinitesimal tools. 
The basic ideas and notations for the coarse-grained
formulation are given in the short \S \ref{CG}.

The main results of the paper are found in \S\S \ref{VREP} -- \ref{collinear}. 
The coarse-grained versions of the Lieb internal energy 
functional $F[\bm{\rho}]$ and the set-valued function 
$\Potl(\bm{\rho})$ which gives the representing potentials 
are the main objects of interest. 
\S \ref{VREP} proves that $F$ is continuous,
that $\Potl(\bm{\rho})$ is non-empty if $\bm{\rho}$
is everywhere non-zero and nowhere spin-saturated,
that $\Potl(\bm{\rho})$ is the functional derivative
of $F$ in a sense appropriate to a general convex
functional, that the map 
$\bm{\rho} \mapsto \{ \bm{\rho}\cdot\bm{v}: 
\bm{v} \in \Potl(\bm{\rho}) \}$ restricted to
nowhere spin-saturated densities is an $L^1$
upper semicontinuous set-valued function and
that the graph of $\Potl$ is closed. 
The reason we have to work with $\Potl$ as a set-valued
function is the well-known non-uniqueness in the SDFT context.
\S\S \ref{U+NU} -- \ref{collinear} deal with this 
problem. There are no nonstandard arguments in these 
sections, so they could probably be read on their own. 
The extent of non-uniqueness in the coarse-grained theory is 
shown to coincide with earlier non-rigorous 
conclusions\cite{Gidopoulos07} for
the continuum theory. Namely, number/spin potentials 
may be non-unique only if the spin-density is saturated
somewhere, or in case of collinear magnetization 
in an eigenstate of $S_z$.
The appropriate conditions are formulated in terms of
densities rather than wavefunctions.
Some concluding remarks are found in \S \ref{conclusion}.
\S\S \ref{U+NU} -- \ref{collinear} have a very different
flavor from that of \S\S \ref{lsc+ns} and \ref{VREP}, but
they are all important parts of a well-rounded picture.

The reader who is curious about the nonstandard arguments might
consider reading quickly through \S \ref{basic-NSA} and then
skipping to \S\S \ref{CG} -- \ref{VREP}.
The reader who just wants to see the results may wish to
start with the summary at the beginning of 
\S \ref{conclusion}, working backwards as needed.

\section{Some basic ideas of SDFT}
\label{basic-DFT}

In this section, we review some basic ideas of nonrelativistic
Density Functional Theory. 
It can safely be skipped by anyone with an acquaintance
with that formalism, after taking note of our notation for spin
densities. The discussion is kept at a heuristic level.

We are concerned with a system of $N$ identical particles
interacting with each other and subject to an external
single-particle potential which functions as a control
parameter. 
The wavefunction of a pure state for this system is a 
function $\psi(z_1,z_2,\ldots,z_N)$ of $N$ positions ${x}_\alpha$ and 
$N$ spin components $s_\alpha$ with respect to some 
quantization axis. These are combined in the abbreviated notation
$z_\alpha = ({x}_\alpha,s_\alpha)$. 
In the usual concrete situations, the particles
are electrons interacting via Coulomb repulsion.
These are spin-$1/2$ fermions, so that the wavefunctions are 
required to be antisymmetric under interchange of 
$z_\alpha$ and $z_\beta$ for $\alpha\not=\beta$.
The value of the particles' spin, and even whether they
are fermions or bosons plays no crucial role.
The discussion in this paper is tailored to the spin-1/2
fermion case, but appropriate modifications can be made for
others.
With the inner product
\beq
\langle \psi | \phi \rangle = \int 
\psi(z)^* \phi(z)\, dz_1\cdots dz_N, 
\eeq
where $\int dz$ denotes integration over position and summation over spin, 
the antisymmetric wavefunctions comprise an $L^2$ Hilbert space 
which will be denoted $\Hilb$.

As alluded to in the introduction, a general Density Functional Theory 
working only with pure states does not get very far. For reasons
of convexity\cite{Lieb83} it is advantageous to allow mixed states,
also called density matrices; this is also a physically reasonable
extension.
Mathematically, a mixed state is represented by a positive trace
class operator in $\Hilb$, having an eigenfunction expansion of the form 
\beq
\gamma = \sum_{i=1}^\infty  c_i |\psi_i \rangle \langle \psi_i |,
\label{mixed-state}
\eeq
in Dirac notation, where $\psi_i$ is an orthonormal set.

The (single-particle) number density is a real function of position,
similarly, the number/spin density is a $2\times 2$ matrix function
of position. For a pure state $\psi$, it is expressed as 
($\alpha$ and $\beta$ on the left-hand side are matrix indices)
\beqa
\rho_{\alpha\beta}({x}) =
N & \int & d{z_2}\cdots d{z_N}\,
\nonumber \\ 
&& \psi({x}\alpha,{z_2},\ldots,{z_N})^*
\psi({x}\beta,{z_2},\ldots,{z_N}).
\eeqa
This is a $2\times 2$ matrix-valued function of position.
With $\sigma_0$ the $2\times 2$ unit matrix and $\sigma_i$, $i=1,2,3$
the Pauli matrices, the number/spin density can be written as
\beq
\rho = \rho_0 \sigma_0 + \rho_1 \sigma_1 + \rho_2 \sigma_2 + \rho_3 \sigma_3
 = \sum \frac{1}{2} \Tr (\sigma_i \rho) \sigma_i.
\eeq
It is therefore convenient to express the number/spin density as the four-vector
\beq
{\rho} \defeq (\rho_0,\vec{\rho}) = (\rho_0,{\rho_1},\rho_2,\rho_3),
\eeq
and refer to it as the {\it 4-density}, the first component being number
density and last three proportional to the spin density. An alternative
notation $(n,2\vec{m}) = (\rho_0,\vec{\rho})$ will also be used, mostly
in \S\S \ref{U+NU} and \ref{collinear}. The factor of $2$ means that
$\vec{m}$ actually integrates to the spin expectation in units of $\hbar$.
The 4-density satisfies $|\vec{\rho}| \le \rho_0/2$ everywhere.
Notation like ``$\psi \mapsto {\rho}$'' is customarily used to indicate
that $\psi$ gives the 4-density $\rho$, but this is sometimes
awkward, so we give the map a name. The 4-density corresponding
to $\psi$ is denoted $\Dns\, \psi$.
This state-to-density map extends additively to mixed states; 
that is, 
\beq
\Dns\, \gamma = \sum_i  c_i \, \Dns\, \psi_i.
\nonumber
\eeq

A physically normalized pure state satisfies $\|\psi\| = 1$, so that
$\int {\rho}_0 \, d{x} = N$. 
Constantly making sure of the normalization is a distracting and
unneccessary nuisance. Thus, we {\em do not} generally assume or 
insist that states and densities be physically normalized,
unless otherwise noted. 

It is traditional to work with the ingredients of the Hamiltonian 
as Hilbert space operators, but quadratic forms are mathematically
convenient and arguably more physically meaningful.
Define the kinetic energy quadratic form by
\beq
\E^K(\psi,\phi) \defeq \frac{1}{2}\int \nabla \psi(\underline{z})^*\cdot 
\nabla \phi(\underline{z}) \, d\underline{z},
\label{KE-Q}
\eeq
the Coulomb interaction energy quadratic form by
\beq
\E^C(\psi,\phi) \defeq \sum_{\alpha<\beta}\int \psi(\underline{z})^* 
\frac{1}{|x_\alpha - x_\beta|} \phi(\underline{z}) \, d\underline{z},
\label{Coulomb-Q}
\eeq
and the total internal energy by
\beq
\E(\psi,\phi) \defeq \E^K(\psi,\phi) + \E^C(\psi,\phi).
\label{E-Q}
\eeq
For the moment, we ignore domain questions.
The abbreviation $\E(\psi,\psi) = \E(\psi)$ is used for diagonal 
elements, and a similar notation is used for $\E^K$ and $\E^C$.

The Lieb internal energy functional\cite{Levy79,Levy82,Lieb83}
for SDFT is defined by
\beq
F[{\rho}] \defeq \inf \{\E(\gamma) : \gamma \in \States, \Dns\, \gamma = {\rho} \}.
\eeq
This is the minimum internal energy consistent with density ${\rho}$.
The Lieb functional is simple and natural in retrospect, but took a long time
to emerge. It solved the original ``V-representability problem'', by extending
the Hohenberg-Kohn internal energy functional to densities regardless
of V-representability (indeed, without any mention of potentials).
To discuss the continuity of $F$, this most central object of DFT,
requires a topology.  One which suggests itself is the $L^1$ norm 
topology which is implicit in the very concept of density.
In other words, we view the 4-densities as a subset of 
$X \defeq L^1({\mathbb R}^3;{\mathbb C^4})$
with the norm 
\beq
\|{f}\| = \int ( |{f}_0(x)|+|\vec{f}(x)| ) \, dx.
\eeq
Then, $F$ is extended to all of $X$ with value
$+\infty$ off the range of the $\Dns$ map; 
this is just a convenience with no physical significance.
There are highly oscillatory densities 
in the range of $\Dns$ for which $F = +\infty$, as well. 

Because the Lieb functional is defined in terms of mixed states
rather than pure states, it is easy to see that $F$ is convex:
\beq
F[\lambda {\rho} + (1-\lambda){\rho}^\prime] 
\le \lambda F[{\rho}] + (1-\lambda)F[{\rho}^\prime],
\label{F-cvx-1}
\eeq
for $0 \le \lambda \le 1$.
This is an extremely important property and a nice way to rephrase it
is in terms of the epigraph of $F$,
$$
{\rm epi}\, F \defeq 
\left\{ ({\rho}^\prime, y) : y \ge  F[{\rho}^\prime]\right\},
$$
which is the region on or above the graph of $F$ in $X \times {\mathbb R}$.
Thus, $\mathrm{epi}\, F$ is a convex set. It is also closed in
the $L^1$ topology\cite{Lieb83} (see Cor. \ref{F-lsc}); this property
corresponds to lower semicontinuity of $F$.

A representing potential ${v}$ for ${\rho}$ is a four-component
function defined by the property
\beq
F[{\rho}] + \langle {v},{\rho}\rangle
\le F[{\rho}^\prime] + \langle {v},{\rho}^\prime\rangle
\quad \forall {\rho}^\prime \in X.
\label{representing-v}
\eeq
The notation
$\langle {v},{\rho}\rangle$ means
\beq
\langle {v},{\rho}\rangle \defeq
\int {v}\cdot{\rho}\, dx
= \int \left( {v_0}{\rho_0} + \vec{v}\cdot\vec{\rho}\right) \, dx.
\eeq
Physically, $\vec{\rho}$ may be thought of as a magnetic field,
though its divergence is unconstrained.

It appears at first sight that some abstract results of
convex analysis\cite{ET} can be applied to the situation.
For example, the Hahn-Banach theorem asserts the existence of
a closed hyperplane separating $({\rho},F[{\rho}])$ from
the interior of ${\rm epi}\, F$. Such a hyperplane would be
the geometrical counterpart of a representing potential for
${\rho}$ in $X^*$, the dual space of $X$.
Similarly, there is a theorem (Prop. I.5.3 of \cite{ET}) which says that when 
there is a unique such hyperplane at $({\rho},F[{\rho}])$,
then it is the G\^{a}teaux derivative of $F$ at ${\rho}$, if
$F$ is continuous at ${\rho}$.
Unfortunately, these theorems get no traction whatever because
$F$ is not continuous. In fact, the effective domain of $F$,
$\mathrm{dom}\, F = \{{\rho}\in X: F[{\rho}] < +\infty \}$ has empty interior.
This is easy to see\cite{Lammert10a,Lammert10b}.
Any open ball around any point in $X$ contains elements of 
$X$ which are negative somewhere, hence not even in $\mathrm{dom} \, F$.
In addition, modifying a density by increasing the amplitude of
arbitrarily short-length-scale oscillations can drive
its kinetic energy arbitrarily high, or even to infinity, with
arbitrarily small change in $X$-norm. 
Lieb\cite{Lieb83} took $L^1({\mathbb R}^3) \cap L^3({\mathbb R}^3)$ 
for $X$ instead of $L^1({\mathbb R}^3)$. 
$L^3$ does not contain the full range of $\Dns$,
but it does contain the densities of finite internal energy,
so is acceptable.
This choice of $X$, however, does not solve the problems just 
mentioned.  
Furthermore, there are reasonable potentials which are not even
linear functionals. For instance, a harmonic potential
$|x|^2$ takes value $+\infty$ on some densities. 
Eq. (\ref{representing-v}) makes perfect sense for 
such a case. 

The only general result on V-representability in the
conventional framework is the fact\cite{Lieb83,Lammert10b} 
that the set of densities which are representable by a potential
in $X^*$ is dense in $\mathrm{dom}\, F$ in $X$-norm.
This is a very weak result, and does not appear to be useful.

Given a pair ${\rho}$, ${v}$ satisfying Eq. (\ref{representing-v}),
${v}$ can be regarded as some sort of ``derivative'' of $F$ at
${\rho}$. But what sort? In particular, we would like to know whether 
$\langle -{v}, {\delta\rho}\rangle$ coincides with the
directional derivatives
\beq
F^\prime[{\rho};{\delta\rho}] \defeq 
\lim_{\epsilon \downarrow 0}{1 \over \epsilon} \left( 
F[{\rho} + \epsilon {\delta\rho}] - F[{\rho}] \right).
\eeq
There is no general guarantee of this, as suggested above
in reference to G\^{a}teaux differentiability.

The questions just raised --- which densities are V-representable
(not to mention, just what qualifies as a potential)?
to what extent do representing potentials coincide with
directional derivatives of $F$? --- are basic elements of
an inquiry into the characteristics, potentially pathological,
of the internal energy functional $F$. 
Currently popular computational algorithms are iterative, 
going through successively more refined approximations to
the density. The above questions are thus relevant to justification
of those algorithms because an understanding of the behavior of $F$ on a 
neighborhood of the sought density is so.
Claims that $F$ has such-and-such a property at physical densities
are necessarily inadequate as the unphyical densities also enter
the computation.
The rest of this paper aims at, and obtains, satisfactory answers
to these questions in a coarse-grained framework, which is slightly 
different and less demanding than the conventional fine-grained
one discussed in this section. Formally, everything is very similar, as the changes
amount to a reinterpretation of density and potential.
With a little nonstandard analysis, the answers for the
coarse-grained theory are obtained easily. 
The next section therefore gives a whirlwind tour of the 
basics of nonstandard analysis.

\section{Some basic ideas of NSA}
\label{basic-NSA}

In the 1960's Abraham Robinson\cite{Robinson66} revived the
old idea of infinitesimals using methods from model theory
(a branch of mathematical logic), and thereby creating the
field of Nonstandard Analysis or infinitesimal analysis. 
Although they are a characteristic feature, however,
infinitesimals are far from the only ``ideal'' objects 
Nonstandard Analysis offers for dealing with mathematical
problems.
In this section, I try to prepare the reader with just enough 
of the jargon and basic ideas that the flavor of arguments in 
the rest of the paper may be appreciated. This flavor 
--- once one gets used to it --- is highly intuitive.
For further background, I highly recommend the notes of 
Lindstrom\cite{Lindstrom85}.
For a highly compressed (twelve-page) summary, see \S 
1 of \cite{Cutland83}.
There are also several good books\cite{Goldblatt,Loeb+Hurd,Loeb,Davis,AFHL,Stroyan+Luxemburg},
as well as some resources\cite{Stroyan-Foundations,Henson}
available on the web.
Several papers\cite{Baty08,Yamashita06,Raab04,Almeida04,Nakamura00}
using NSA have appeared in this journal in recent years.

In (model theoretic) nonstandard analysis, there is a nonstandard
counterpart of every conventional mathematical object.
A natural place to start is with the traditional reals, ${\mathbb R}$.
Its counterpart is $\Star{\mathbb R}$, the ordered field of hyperreals,
which is an extension of ${\mathbb R}$:
in addition to the familiar real numbers, $\Star{\mathbb R}$ contains 
infinitesimals and their reciprocals, the illimited hyperreals. 
$\epsilon \in \Star{\mathbb R}$ is infinitesimal if $|\epsilon| < \delta$ 
for every $\delta > 0$ in ${\mathbb R}$, whereas $x$ is illimited 
(called ``infinite'' by many) if $|x| > n$ for every natural
number $n$.  The notation ``$n \approx +\infty$''
($n \approx -\infty$) means that $n$ is positive (negative) illimited,
and $x \ll \infty$ means that $x$ is limited.

Two hyperreals $x$ and $y$ are infinitely close,
$x\approx y$, if their difference is an infinitesimal.
Thus, ``$x$ is infinitesimal'' and ``$x\approx 0$'' are synonymous. 
$x \lesssim y$ means that $x$ does not exceed $y$ by more than
an infinitesimal.

In fact, every object (formally viewed as a set) of
conventional mathematics has a nonstandard enrichment.
The enrichment of the naturals is $\Star{\mathbb N}$, the hypernaturals.
$\Star{\mathbb N}$ contains $\mathbb N$, as well as illimited hypernaturals.

Not only objects, but (first order) mathematical statements have
$\star$-transforms. 
Consider a sequence $(x_n)_{n\in{\mathbb N}}$ of real numbers.
It has a $\star$-transform which is a map from $\Star{\mathbb N}$ 
to $\Star{\mathbb R}$, and which we continue to denote by $x_n$.
The statement ``$\lim_{n\to\infty} x_n = a$'', is equivalent to
`` given $k \in {\mathbb N}$,
there is $m_k$ such that 
$\forall n \in {\mathbb N},\, n > m_k \Rightarrow |x_n - a| < 1/k$''. 
${\mathbb N}$, $a$ and $m_k$ are parameters in this statement. Since
$\Star{a} = a$ and $\Star{m_k} = m_k$, the $\star$-transform
of the statement is 
$\forall n \in \Star{\mathbb N},\, n > m_k \Rightarrow |x_n - a| < 1/k$. 
To obtain the $\star$-transform, just ``put stars on everything''.
The point of the operation is that the {\it Transfer Principle} 
asserts that statement in the standard universe is true if and 
only if its $\star$-transform is true in the nonstandard universe. 
In the case at hand, an illimited $n$ will satisfy all the statements 
as $k$ runs through ${\mathbb N}$, so $x_n \approx a$ for 
illimited $n$. And conversely, if $x_n \approx a$ for illimited
$n$, $a$ is the limit of the original sequence.

In a similar way, one finds the nonstandard characterization
of continuity for a function $f: {\mathbb R} \to {\mathbb R}$.
Namely, for $x \in {\mathbb R}$ and $y\in \Star{\mathbb R}$
with $y \approx x$, $f(y) \approx f(x)$. As just happened with
$\Star{f}(y)$, stars will sometimes be dropped when no ambiguity 
is possible. 
If $x$ is standard and $y \approx x$, then $x$ is
the {\it standard part} of $y$, and we write $x = \Std y$ or
$x = \mathrm{st}\, y$. For a standard point $x$, the collection of
all points infinitely close to it is called its {\it monad}.
Continuity of $f$ at $x$ is equivalent to ``$f$ maps the monad of
$x$ into the monad of $f(x)$''. 
These concepts of monad and standard part generalize to arbitrary
topological spaces.

For another example, consider the least number principle:
\beq
\forall B \in {\mathcal P}({\mathbb N}),\, \exists x \in B :
y \in B \Rightarrow x \le y.
\nonumber
\eeq
This has $\star$-transform
\beq
\forall B \in \Star{\mathcal P}({\mathbb N}),\, \exists x \in B :
y \in B \Rightarrow x \le y.
\nonumber
\eeq
The one free parameter, ${\mathcal P}({\mathbb N})$, the power set of
the naturals, has been replace by $\Star{\mathcal P}({\mathbb N})$.
This is {\em not} the power set of $\Star{\mathbb N}$.
Instead, it is the set of {\it internal} subsets of $\Star{\mathcal N}$.
A set is {\it standard} if it is the $\star$-transform of a
conventional set, for example $\Star{\mathbb N}$. A set is
internal if it is an element of a standard set.
Standard sets are also internal.
Not all subsets of $\Star{\mathbb N}$ are internal.
For example, $\Star{\mathbb N}\setminus{\mathbb N}$ is not.
Indeed, if $n$ is illimited, so is $n-1$. Thus, since the set 
of illimited hypernaturals does not obey the least number 
principle, it must not be internal.
It is {\it external}.  Similar reasoning leads to the conclusion
that the set of infinitesimals in $\Star{\mathbb R}$, or the monad
of any point, is an external set. 

The internal/external dichotomy is subtle, but crucially important. 
If $A$ is an internal subset of $\Star{\mathbb N}$ containing
all illimited $n$, then there must be some limited $m$ such
that $n \ge m \Rightarrow n \in A$. This consequence of the
least number principle is called {\it underflow} (or {\it underspill}).
Similarly, an internal subset of $\Star{\mathbb N}$ containing
all sufficiently large limited numbers contains also some
illimited numbers ({\it overflow} or {\it overspill}), and 
an internal subset of $\Star{\mathbb R}$ containing all non-negative
infinitesimals contains all $x < \epsilon$ for some $\epsilon > 0$ 
in ${\mathbb R}$.

How can we recognize internal sets, apart from the definition? 
The {\it Internal Definition Principle} gives an answer.
If $A$ is an internal set and $\phi(x;A,B_1,\ldots,B_n)$ is
a formula with internal parameters $A, B_1,\ldots,  B_n$, then
the subset of $A$ consisting of the elements satisfying $\phi$,
$\{ x\in A: \phi(x,A,B_1,\ldots,B_n)\}$ is also internal.
Suppose an internal sequence $(x_n)_{n\in \Star{\mathbb N}}$
is given with the property that $x_n \approx 0$ for all
limited $n$. Then, the set $\{ m \in \Star{\mathbb N} :
|x_m| < 1/m \}$ is internal. The parameters in it are
$\Star{\mathbb N}$ and $(x_n)$, both of which are internal.
Thus, that set is internal by the Internal Definition Principle,
and therefore contains all $m$ up to some
$N$ in $\Star{\mathbb N} \setminus {\mathbb N}$, implying that
$x_m \approx 0$ for all $m \le N$. (``Robinson's Lemma'')

Another important principle is {\it Saturation}.
For an infinite cardinal $\kappa$, a $\kappa$-saturated 
nonstandard model has the property that, if 
$\{A_j\}_{j\in J}$ is a collection of internal sets
with an index set of cardinality less than $\kappa$ 
and the finite intersection property, then
$\cap_{j\in J} A_j \not= \emptyset$. Usually 
$\aleph_1$-saturation is adequate for applications.
We assume that much, at least.

The sequence space $\ell_1$ and its nonstandard enrichment
is a prototype of some of the spaces
which will be used later. By definition, 
$\ell_1$ is the Banach space of sequences $x: {\mathbb N} \to {\mathbb C}$
with finite norm $\| x \| = \sum_{j=1}^\infty |x_j| < \infty$.
Then, $\Star\ell_1$ consists of sequences
$\Star{\mathbb N} \to \Star{\mathbb C}$ with norm
$\| y \| = \sum_{j\in\Star{\mathbb N}} |y_j|$ mapping
$\Star\ell_1$ into $\Star{\mathbb R}$. If $x$ is standard,
(in $\ell_1$) then $y$ is in the monad of $x$ ($y \approx x$)
if 
$\| y - x \| = \sum_{j\in\Star{\mathbb N}} |y_j-x_j| \approx 0$.
Here, the $\star$ has been dropped on the nonstandard extensions 
of both the standard element $x$ and the norm. 
We seek a more explicit condition ensuring $y \approx x$.

$y$ is said to be {\it nearstandard} if it is in the monad of
some standard vector in $\ell_1$, which must therefore be
its standard part $\Std y$.
I claim that $y$ is nearstandard precisely when
$\| y\|$ is limited and $y$ ``has infinitesimal tail'':
$\sum_{i\ge n} |y_i| \approx 0$ for all illimited $n$. 
And, in that case the standard part of $y$ is given by
\beq
(\Std y)_i = \Std(y_i), \quad \text{for limited }\, i.
\label{std-part-limited-i}
\eeq
The first condition is obvious. The second follows from the
fact that a standard $x\in \ell_1$ has infinitesimal tail:
given $k \in {\mathbb N}$, there is $N_k$ such that
$n \ge N_k \Rightarrow \sum_{i\ge n} |x_i| < 1/k$. By Transfer, 
$n \approx +\infty$ satisfies all of these statements
as $k$ varies, and therefore $\sum_{i\ge n} |x_i| \approx 0$. 
(The $\star$ on `$\Star{x}$' has been dropped again.)
So, $y \approx x$ implies $y$ also has infinitesimal tail.
All that remains to verify, then is that (\ref{std-part-limited-i})
gives the standard part. Define $x \in \ell_1$ by
$x_i = \Std(y_i)$ for limited $i$.
Then, given any natural $k$, the set $A_k = \{ n: \sum_{i=1}^n |y_i - x_i| < 1/k\}$
contains all $n\in {\mathbb N}$. It is also internal, since it has two
standard parameters ($k$ and $x$) and one internal but nonstandard parameter
($y$). Thus, there is an illimited $N$ in $A_k$.
But then, $\sum_{i\in\Star{\mathbb N}} |y_i - x_i| = 
\sum_{i \le N} |y_i - x_i| + \sum_{i > N} |y_i - x_i| \lesssim 1/k$
because $N \in A_k$ and both $y$ and $x$ have infinitesimal tail.
This being true for every $k$, $\| y - x \| \approx 0$ and
$x$ is $\Std y$, the standard part of $y$.

The nonstandard characterizations of open, closed and compact
subsets of a topological space are simple and useful.
${\mathrm{st}}^{-1} A$ denotes the union of the monads of all
points of $A$. 
$A$ is open if ${\mathrm{st}}^{-1} A \subset \Star{A}$, it is
compact if $\Star{A}  \subset {\mathrm{st}}^{-1} A$, and it is
closed if ${\mathrm{st}}^{-1} A$ contains all nearstandard points
in $\Star{A}$.

\section{Lower semicontinuity and nearstandard mixed states}
\label{lsc+ns}

This Section establishes lower semicontinuity of the energy forms
introduced in \S \ref{basic-DFT}, both for pure states and mixed states.
Some of the consequences, particularly Corollaries \ref{minimizer-existence} 
and \ref{F-lsc}, will be important later. 
The results of Lemmas \ref{E1-lsc-on-H} and \ref{E1-lsc-on-S} are 
well-known (see, e.g. \cite{Simon77}), but the nonstandard viewpoint taken
here is novel.

We begin with a nonstandard approach to lower semicontinuity of $\E$
on $\Hilb$, which is a case of the equivalence of closedness and lower 
semicontinuity of positive quadratic forms\cite{Simon77}.
Returning to the quadratic forms (\ref{KE-Q},\ref{Coulomb-Q},\ref{E-Q}),
regard them as initially defined on the space of (antisymmetric)
${\mathbb C}^{2N}$-valued infinitely differentiable functions of 
rapid decrease at infinity (the components correspond to spin indices).
That is,
$(1+|\underline{z}|^2)^m \partial^\alpha f/\partial z^\alpha$ is
bounded for all $m$ and multi-indices $\alpha$.
Denote this ``seed space'' by $S$;  other choices work, but this one has 
especially nice properties with respect to Fourier transformation.

It is very important that the form $\E^C$ is bounded relative to $\E^K$ 
with relative bound less than $1$\cite{RSII,Teschl,Hislop+Sigal},
that is, for some $0 < a < 1$,  a sufficiently large $b$ can
be found such that
\beq
\E^C(\psi) \le a \E^K(\psi) + b \|\psi\|^2.
\label{relative-bddness}
\eeq
This means that the non-degenerate inner product
\beq
\langle x|y \rangle_{\E} = \E(x,y) + \langle x | y \rangle
\eeq
is equivalent to $\langle x|y \rangle_{K}$, defined similarly
but with $\E$ replaced by $\E^K$. That means that there is
some $c > 1$ satisfying
$c^{-1} \langle \cdot | \cdot \rangle_K \le
\langle \cdot | \cdot \rangle_\E \le
c \langle \cdot | \cdot \rangle_K$. 
One nonstandard way to construct the completion of $S$ with respect to 
$\langle \cdot | \cdot \rangle$ or $\langle \cdot | \cdot \rangle_K$
is as a weak nonstandard hull. This abstract construction 
actually applies to completion of any linear space $S$ with respect to 
a non-degenerate inner product. Let $W$ be the collection of seminorms
$x \mapsto |\langle x | \phi \rangle|$ for $\phi \in S$, initially
defined as real-valued on $S$, but immediately extended to hyperreal
seminorms on $\Star{S}$. Now define the set of $W$-infinitesimal elements 
$\mathrm{inf}_W \Star{S}$ of $\Star{S}$ to be those satisfying 
$|\langle x | \phi\rangle| \approx 0$ for every $\phi \in S$, and the
$W$-finite elements $\mathrm{fin}_W \Star{S}$ those satisfying
$|\langle x | \phi\rangle| \ll \infty$ for every $\phi \in S$.
The completion of $S$ is obtained as the quotient
$\hat{S} = \mathrm{fin}_W \Star{S}/\mathrm{inf}_W \Star{S}$,
and the norm is $\|\hat{x}\| = \sup \left\{\Std |\langle x | \phi \rangle| :
\phi \in S, \|\phi\| = 1\right\}$.
$S$ can be viewed as a subset of $\hat{S}$ by
identifying elements with their equivalence classes. To see that $\hat{S}$ is 
complete, consider a Cauchy sequence $(\hat{x}_n)$ in $\hat{S}$,
lifting to a sequence $(x_n)$ in $\Star{S}$. By overspill and
countable saturation, the sequence can be extended up to
$N \in \Star{\mathbb N}\setminus{\mathbb N}$ still respecting
the Cauchy condition
$m > k_n \Rightarrow \| \hat{x}_m - \hat{x}_{k_n} \| < 1/n$ for
each $n$ in ${\mathbb N}$. Thus, $\hat{x}_N$ is the sought limit of
the sequence in $\hat{S}$. 

Applying this construction using the $L^2$ inner product to obtain $\hat{S}$,
and the $\E^K$ inner product to obtain $\hat{S^K}$, it becomes easy to see
that the latter can be identified with a subspace of $\hat{S}$.
For, since $\langle x | \phi \rangle_K = \langle x | H_{00} \phi \rangle$,
where $H_{00}$ is the kinetic energy Hamiltonian (well-defined as a 
map $S \to S$),
$\langle \cdot | \cdot \rangle$ monads are no larger than
$\langle \cdot | \cdot \rangle_K$ monads. On the other hand,
they are also obviously no smaller due to the relative weakness
of the $\|\cdot \|$ norm compared to the $\|\cdot\|_K$ norm. 

On $\Hilb$, the ordinary $L^2$ and $\E^K$ norms can be found according
to $\| x \| = f(x)$, $\| x \|_K = f_K(x)$, where
$f(x) \defeq 
\mathrm{sup} \{ |\langle \phi | x \rangle | : \phi \in S, \|\phi \| = 1 \} $,
and $f_K(x) \defeq 
\mathrm{sup} \{ |\langle \phi | x \rangle_K | : \phi \in S, \|\phi \|_K = 1 \}$. 
These take values in $[0,\infty]$.
Since each seminorm $x \mapsto |\langle \phi | x \rangle_K | =
|\langle H_{00} \phi | x \rangle |$ is weakly continuous, the 
supremum is lower semicontinuous. Another way to this conclusion is
to look at the function $\Star{f}_K$ on $\Star\Hilb$:
$\Star{f}_K(x) = 
\mathrm{sup} \{ |\langle \phi | x \rangle_K | : \phi \in \Star{S}, \|\phi \|_K = 1 \}$. 
Now, if $x$ is standard, that is to say, in $\Hilb$, then $\Star{f}(x) = f(x)$.
For $y \approx x$,  
$|\langle \phi | x \rangle_K | \approx |\langle \phi | y \rangle_K |$
for every $\phi \in S$, yet the supremum defining $\|y\|_K$ includes
also $\phi \in \Star{S}\setminus{S}$, so that $\|y \|_K \gtrsim \|x\|_K$
which is exactly the nonstandard characterization of lower semicontinuity.

\begin{lem}
\label{E1-lsc-on-H} 
$\E_1$ is lower semicontinuous on ${\Hilb}$
\end{lem}
\begin{proof}
The proof is contained in the preceding two paragraphs.
Also, see Reference \cite{Simon77}.
\end{proof}

Turning to mixed states, $\Gamma \in {}^\star\States$ is nearstandard
if it is nearstandard with respect to trace-norm (the only topology
we consider on $\States$).
It will be useful to characterize this more concretely in terms
of nearstandardness in $^\star\Hilb$. 
Whenever possible and profitable we try to reduce properties of mixed 
states to corresponding, but more intuitively graspable properties of pure states.
Now, note that for 
$\gamma = \sum_{i \in {\mathbb N}}  c_i |\psi_i \rangle \langle \psi_i |$ 
in $\States$, 
given $\epsilon > 0$ in ${\mathbb R}$, there is a finite-rank mixed state
within $\epsilon$ of $\gamma$ in trace norm. Indeed, some truncation 
$\sum_{i = 1}^m  c_i |\psi_i \rangle \langle \psi_i |$ will serve
the purpose. Otherwise put, the finite-rank mixed states are dense 
in $\States$. Thus, a mixed state
$\gamma = \sum_{i \in {}^\star {\mathbb N}} 
 c_i |\psi_i \rangle \langle \psi_i |$ in ${}^\star\States$ is
nearstandard if and only if it can be approximated to any given
standard accuracy by some partial sum
$\sum_{i \in J} {}^\circ c_i |{}^\circ\psi_i \rangle \langle {}^\circ \psi_i |$
with $|J| < \infty$. With the terms of the sum ordered by decreasing eigenvalue,
this is equivalent to: $\gamma_i$ is nearstandard whenever $ c_i \not\approx 0$,
and $\sum_{i=1}^\infty {}^\circ{c_i} = {}^\circ \Tr \gamma$.
Also, note that trace-norm topology is stronger than uniform topology, which in
turn is stronger than strong-operator topology, so that
$\gamma \approx \gamma^\prime$ only if every eigenvector of $\gamma$ is
infinitely close to being an eigenector of $\gamma^\prime$ with
the same eigenvalue: if $(\gamma - a)\psi = 0$, then $(\gamma^\prime - a)\psi \approx 0$.

The way is now prepared for the analog of Lemma \ref{E1-lsc-on-H} for mixed states.
\begin{lem}
\label{E1-lsc-on-S} 
$\E_1$ is lower semicontinuous on ${\States}$
\end{lem}
\begin{proof}
Take $\gamma \in {\States}$, with nonstandard enrichment
$
{}^\star\gamma = \sum_{i \in {}^\star {\mathbb N}} 
 c_i |\psi_i \rangle \langle \psi_i |,
$
and suppose 
$\gamma^\prime =
\sum_{i \in {}^\star {\mathbb N}} 
 c_i^\prime |\psi_i^\prime \rangle \langle \psi_i^\prime |$
satisfies $\gamma^\prime \approx {}^\star\gamma$.
We need to show that $\E_1(\gamma^\prime) \gtrsim \E_1(\gamma)$.
As just discussed, every eigenvector $\psi_i^\prime$ of 
$\gamma^\prime$ is infinitely close to an eigenvector $\psi_i$ of 
$\gamma$ with infinitely close eigenvalue, and {\it vice versa}.
But, since the eigenvectors of $\gamma$ are standard, Lemma \ref{E1-lsc-on-H} 
implies that
\beq
\E_1(\psi_i^\prime) \gtrsim \E_1({}^\circ\psi_i^\prime) = \E_1(\psi_i).
\nonumber
\eeq
(We assume that the terms are arranged in decreasing order of
eigenvectors and in case of equal eigenvalues, appropriate choices 
of basis must be made.) Thus,
$$
\sum_{i=1}^m  c_i^\prime \E_1(\psi_i^\prime) 
\gtrsim \sum_{i=1}^m  c_i \E_1(\psi_i),
$$ 
for any $m \in {\mathbb N}$.
From this, we immediately
conclude that $\E_1(\gamma^\prime) \gtrsim \E_1(\gamma)$.
\end{proof}

The map $\Dns$ from $\Hilb$ to $L^1$ is continuous, therefore
if $\psi \in {}^\star\Hilb$ is nearstandard, $\Dns\, \psi$ is
also $L^1$-nearstandard. Going the other way, a weaker condition 
on $\Dns\, \psi$, together with energy-boundedness of $\psi$
suffices to guarantee nearstandardness of $\Psi$ as we shall see.

If for any $\epsilon > 0$, there is some $R$ such that
$\int_{{x} \ge R} \rho\, d{x} < \epsilon$, we will
say that $\rho$ is ``nearstandardly concentrated''.
Note that, by underspill this is equivalent to
$\int_{{x} \ge R} {}^\star\rho\, d{x} \approx 0$ 
for all illimited $R$.

\begin{lem}
\label{ns-pure-states}
If $\Psi \in {}^\star\Hilb$ has nearstandardly concentrated density
and $\E_1(\Psi)$ is limited, then $\Psi$ is nearstandard.
\end{lem}
\begin{proof}
Note that, since $\Hilb$ is a {\em complete} metric space, it suffices to 
show that $\Psi$ can be approximated to any standard accuracy by a vector
in $\Hilb$.
Also, by relative form boundedness of $\E$ with respect to $\E^K$,
together with positivity of $\E^C$, there is some constant
$c$ such that $c^{-1} \E_1 < \E_1^K < c \E_1$, so that $\E$
and $\E^K$ are interchangeable for our purposes here. 

The first step is to show that $\Psi$ can be approximated by 
a ${}^\star\Hilb$ vector with bounded support and finite energy
(a ``smooth truncation''). Then we will be able to assume that $\Psi$
itself is such a vector with no loss of generality.
So, let $\chi({x})$ be a smooth cutoff function which is nonincreasing
with $\|{x}\|$, equal to one in $B_{1}$ and supported in $B_2$, where
$B_R$ is the ball $\|{x}\| \le R$.
For $R > 0$, define the $N$-particle cutoff function by
\beq
\hat{\chi}_R = \prod_{\alpha=1}^N \chi({x}_\alpha/R),
\nonumber 
\eeq
and cutoff wavefunction by
\beq
\Phi_R = \hat{\chi}_R \Psi.
\nonumber
\eeq
Thus, $\Phi_R$ has zero probability for any particle to be outside
the $2R$-ball. By the assumption of nearstandardly concentrated density, 
$\|\Phi_R - \Psi \|$ can be made smaller than any standard tolerance
by taking $R$ large enough. Further, since $\nabla \Phi_R = 
\hat{\chi}_R \nabla \Psi + (\nabla \hat{\chi}_R) \Psi$
and $\hat\chi_R$ is smooth, $\E_1^K(\Phi_R)$ is limited.
Thus, we may assume that $\Psi$ has bounded support,
contained in the closed box $\overline{\Lambda}_L = \{|x|,|y|,|z| \le L\}$,
and we now consider that case.

Using the particle-in-a-box wavefunctions for $\overline{\Lambda}_L$, we can 
construct an explicit orthonormal basis for $\Hilb\upharpoonright_L$, 
the closed subspace of $\Hilb$ consisting of vectors with density
supported in $\overline{\Lambda}_L$. For any $E$, there are a finite number of
basis vectors with $\E^K < E$. Thus, given any $E$ and $\epsilon$,
there is a finite orthonormal set $\varphi_1,\ldots,\varphi_n$
such that whenever $\phi \in \Hilb\upharpoonright_L$ satisfies
$\E_1(\phi) < E$, then the component of $\phi$ orthogonal  
to $\mathrm{span}(\varphi_1,\ldots,\varphi_n)$ has 
norm less than $\epsilon$. 
By Transfer, this continues to hold for 
$\phi \in {}^\star\Hilb\upharpoonright_L$. But, the projection
of $\phi$ onto $\mathrm{span}(\varphi_1,\ldots,\varphi_n)$ 
is clearly nearstandard. 
Since $E$ and $\epsilon$ are free, this means that $\Psi$
can be approximated to any standard tolerance by a nearstandard
vector, and therefore is itself nearstandard. 
\end{proof}

Here is the central result of this Section.
\begin{prop}
\label{ns-DM}
If, $\Gamma\in {}^\star\States$ has
nearstandardly concentrated density
and $\E_1(\Gamma)$ is limited, then $\Gamma$ is nearstandard.
\end{prop}
\begin{proof}
Let $\Gamma$ have eigenfunction expansion
\beq
\Gamma = \sum_{i \in {}^\star {\mathbb N}} 
 c_i |\psi_i \rangle \langle \psi_i |.
\nonumber
\eeq
Since $\Dns$ and $\E_1$ are additive and positive maps, 
whenever $ c_i$ is not infinitesimal, $\psi_i$
is nearstandardly concentrated and of limited energy,
hence nearstandard by Lemma \ref{ns-pure-states}.
Thus, with 
\beq
 \Gamma^\prime \defeq \sum_{i=1}^\infty {}^\circ c_i 
| {}^\circ\psi_i\rangle \langle {}^\circ\psi_i |,
\nonumber
\eeq
it follows that ${}^\circ\Gamma = \Gamma^\prime$ 
unless the sum of infinitesimal weights $ c_i$ is noninfinitesimal,
that is $\Tr \Gamma^\prime < {}^\star\Tr \Gamma - \epsilon$, for
some $\epsilon > 0$ in ${\mathbb R}$. 
In that case, defining
$J_n \defeq \{ i:  c_i < 1/n \}$, the {\em internal} set 
$K \defeq \{ n\in {}^\star{\mathbb N}: \sum_{J_n}  c_i > \epsilon \}$ 
contains all of ${\mathbb N}$, hence some illimited $\omega$,
and $|J_\omega| > \epsilon \omega$.

Now define the mixed state
\beq
\tilde{\Gamma} \defeq \sum_{i\in J_\omega}  c_i |\psi_i \rangle \langle \psi_i |
< \Gamma.
\nonumber
\eeq
Since $\Hilb$ is separable, fewer than $\epsilon \omega/2$ of the
(orthogonal) eigenvectors $\psi_i$ of $\tilde{\Gamma}$ can be nearstandard, and
they carry a combined weight not exceeding $(\epsilon\omega/2)(1/\omega) = \epsilon/2$.
The remaining eigenvectors are not nearstandard, so each must have either
illimited $\E$, or non-nearstandardly concentrated density, or both. 
But since these non-nearstandard vectors
enter $\tilde{\Gamma}$ with a combined weight of at least $\epsilon/2$, 
$\tilde{\Gamma}$ itself and {\it a fortiori} $\Gamma$ must also have
either illimited $\E$ or non-nearstandardly concentrated density.
But, that is contrary to hypothesis.
\end{proof}

This Proposition has a couple of easy but important corollaries.

\begin{cor}
\label{minimizer-existence}
If ${\rho} \in X$ satisfies $F[{\rho}] < \infty$, then 
there exists $\gamma \in \States$ with $\Dns\, \gamma = {\rho}$ and 
$\E(\gamma) = F[{\rho}]$.
\end{cor}
\begin{proof}
Since $F[{\rho}] < \infty$, there is some $\gamma \in {}^\star\States$ with
$\Dns\, \gamma = {\rho}$ and $\E(\gamma) \approx F[{\rho}]$.
By Prop. \ref{ns-DM}, 
$\gamma$ is nearstandard. 
Then, since $\E$ is lower semicontinuous, bu
Lemma \ref{E1-lsc-on-S},
$\E(\rm{st} \, \gamma) \lesssim F[{\rho}]$.
Since both sides of this inequality are standard, 
\beq
\E(\rm{st} \, \gamma) \le F[{\rho}].
\nonumber
\eeq
On the other hand, $\Dns : \States \to X$ is 
continuous, which implies that ${\rho}$ is
in the monad of $\Dns \, \rm{st}\, \gamma$.
But, ${\rho}$ is standard, so $\Dns\, \rm{st}\, \gamma = {\rho}$.
Therefore, $\E(\rm{st}\, \gamma) \ge F[{\rho}]$, which combined 
with the previous display yields $\E(\rm{st}\, \gamma) = F[{\rho}]$.
\end{proof}

\begin{cor}
\label{F-lsc}
Let $\mathcal T$ be a Hausdorff topology on $\mathrm{Range}\, \Dns$
which makes the map $\Dns\,$ from mixed-states to 4-densities continuous. 
then $F$ is lower semicontinuous with respect to ${\mathcal T}$.
\end{cor}
\begin{proof}
Let ${\rho}$ be a 4-density in $\mathrm{Range}\, \Dns$
and ${\rho}^\prime \in 
\Star\mathrm{Range}\, \Dns = \mathrm{Range}\, \Star\Dns$
such that ${\rho}^\prime \approx {\rho}$.
And, let $\gamma^\prime$ be a ${}^\star$density matrix with 
$\Dns\, \gamma^\prime = {\rho}^\prime$.
By Prop. \ref{ns-DM}, $\gamma^\prime$ is nearstandard. 
Thus, $\Dns\, \Std \gamma^\prime$ is a standard density which
is infinitely close to ${\rho}^\prime$ by the assumption of
continuity of $\Dns$.
Therefore, $\Dns\, \Std \gamma^\prime = {}^\circ {\rho}^\prime = {\rho}$;
the Hausdorff assumption is needed here to ensure that the standard part
operation is well defined.
Then,
$$
F[{\rho}^\prime] = \E_1(\gamma^\prime) \gtrsim
\E_1({}^\circ \gamma^\prime) \ge F[{\rho}],
$$
where the almost-inequality follows from Lemma \ref{E1-lsc-on-S} 
and the final inequality by definition of $F[{\rho}]$.
\end{proof}

\section{coarse-graining}
\label{CG}

This section reviews the basic notions of the coarse-grained DFT
framework\cite{Lammert06,Lammert10a,Lammert10b}, augmented to
handle SDFT. Much of it looks nearly the same on the surface
as the conventional fine-grained theory of \S \ref{basic-DFT}.
Building on the results of \S \ref{lsc+ns}, the existence
and regularity results for the coarse-grained theory will be
presented in \S \ref{VREP}. In addition to the considerably
simpler and more intuitive proofs, those results go beyond
the previous version of the theory\cite{Lammert06,Lammert10a,Lammert10b} 
by treating spin-densities.

\subsection{densities}

Begin by partitioning ${\mathbb R}^3$ into a regular
array of cubical cells $\Omega_i$ of side length $\ell$;
the partition is denoted by $\Part$.
A coarse-grained 4-density is simply a specification of the average 
4-density in each cell. 
As such, a coarse-grained 4-density is an equivalence class
of densities in $\X$, the members of the class differing only 
in the way particle number and spin is distributed {\it within} the cells.
Generally, coarse-grained densities are denoted by the same sorts of
symbols as were used previously for fine-grained densities, but
there should be no occasion for confusion as the rest of the paper focusses on
coarse-grained densities. As described above, the coarse-grained density 
${\bm \rho}$ as simply a list of the net spin and particle number in 
each cell (divided by $\ell^3$). 
There is nothing wrong with that, but it is more convenient and intuitive 
to think of coarse-graining as a {\it levelling} 
(or {\it projection}, see below): without altering the 
net number and spin in any cell, we level it out so that it is 
uniform across each cell. Then we identify ${\bm \rho}$ with the resulting
cell-wise constant function. The discontinuities of this {\it representation} 
of $\bm{\rho}$ have no significance.
Note that it also makes no difference to which cell the (shared) faces are
assigned since they have zero Lebesgue measure.
The coarse-grained densities, thus understood, belong to the $L^1$ space
of cell-constant 4-component functions with norm
\beq
\| {\bm f} \|_1 \defeq \int_{{\mathbb R}^3} 
\left( |f_0({x})| + |\vec{f}({x})|\right) \, d{x}.
\eeq
This Banach space, which will be denoted $\X$, is a subspace of
$X$, so we could also describe matters with the aid of the 
projection 
\beq
\pi: X \to \X
\eeq
which averages over cells of $\Part$.

Some useful subspaces of $\X$ are singled out:
\beqa 
\X^+ &\defeq & \left\{ \bm{\rho} \in \X: \rho_0 \ge 0,
\, \mathrm{and}\, |\vec{\rho}| \le \rho_0 \right\}
\nonumber \\
\X^{++} &\defeq & \left\{ \bm{\rho} \in \X^+: \rho_0 > 0 
\right\}
\eeqa
$\X^+$ consists of true densities, $\X^{++}$ adds the restriction
that the number-density does not vanish on any cell.
If $|\vec{\rho}| = \rho_0$ on some cell, that
cell is said to be {\it spin-saturated}. Subspaces without
spin saturation
\beqa
\X^\oplus &\defeq & \left\{ \bm{\rho} \in \X^+: 
|\vec{\rho}| < \rho_0 \,\mathrm{where}\, \rho_0 > 0 \right\}
\nonumber \\
\X^{\oplus\oplus} &\defeq & \left\{ \bm{\rho}  \in \X^{++}:
|\vec{\rho}| < \rho_0 \right\}
\label{unsaturated-subspaces}
\eeqa 
are also defined.  

???????
Only  $\X^{\oplus\oplus}$ can be fully controlled in a general way,
but collinear spin-saturated states will be treated in \S \ref{collinear}.

A fine-grained density in the conventional theory  
has associated with it a whole set of states, and some
are selected out by an energetic criterion in the definition
of the Lieb internal energy functional. For the remainder of the
paper, we write that fine-grained internal energy functional
of \S \ref{basic-DFT} as $\hat{F}$. Henceforth, $F$ will denote its
coarse-grained analog defined for $\bm{\rho} \in \X$ by
\begin{align}
F[{\bm \rho}] & \defeq 
\inf \left\{
{\hat{F}}[\alpha] : \, \alpha\in X, \, \pi \alpha = \bm{\rho} \right\} 
\nonumber \\
& =
\inf \left\{
{\E}(\Gamma) : \, {\Gamma\in\States},\, \pi\, \Dns\, \Gamma = {\bm \rho} 
\right\},
\label{FL}
\end{align}
in perfect analogy to the conventional case.
If there are no states with $\pi \Dns\, \Gamma = {\bm \rho}$,
then $F[{\bm\rho}] = +\infty$, as usual.
But, there are states realizing any density in $\X^+$,
and 
\beq
F[{\bm\rho}] \le \|\bm{\rho}\|_1 V_0, \quad \mathrm{for}\, 
\bm{\rho} \in \X^+,
\label{F-bound}
\eeq
where the constant $V_0$ is the one given in (\ref{F-bound}).
This is in stark contrast to the continuum situation, and 
arises because there is a finite internal energy cost $N V_0$
to putting all the particles in the same cell of $\Part$.
Any density in $\X^+$ can at least be realized by a 
mixed-state sum of such single-cell states, and the
bound follows.
Since for $t\in [0,1]$, the
convex combination $t \Gamma + (1-t) \Gamma^\prime$ 
of density matrices gives the corresponding convex combination 
of densities, it is immediate that $F$ is convex:
\beq
F[t {\bm \rho} + (1-t) {\bm \rho^\prime}] 
\le t F[{\bm \rho}] + (1-t) F[{\bm \rho^\prime}].  
\label{F-cvx}
\eeq

The coarse-grained internal energy functional shares important
properties with its fine-grained analog, such as existence of
a minimizing mixed state in the definition (\ref{FL}), and 
lower semicontinuity. Both of these are easily established
by appeal to Prop. \ref{ns-DM}.
Given $\bm{\rho} \in \X^+$ (standard), find 
$\gamma \in \Dns^{-1} \pi^{-1}\bm{\rho}$ such that
$\E[\gamma] \approx \hat{F}[\bm{\rho}]$ Then, taking standard parts
and using continuity of $\Dns$ and $\pi$,
$\pi \Dns \Std{\gamma} = \Std{\bm{\rho}} = \bm{\rho}$, and
$\E[\Std{\gamma}] \lesssim F[\bm{\rho}]$. But, since both sides
of this near-inequality are standard, it must be `$\le$'.
For lower semicontinuity, take $\bm{\rho^\prime} \approx \bm{\rho}$, and 
$\alpha^\prime \in \pi^{-1} \bm{\rho^\prime}$ such that
$\hat{F}[\alpha^\prime] \approx F[\bm{\rho^\prime}]$.
Then, again $\hat{F}[\Std{\alpha^\prime}] \lesssim \hat{F}[\alpha^\prime]$,
but $\pi (\Std{\alpha^\prime})
= \Std(\pi {\alpha^\prime}) = \Std{\bm{\rho^\prime}} = \bm{\rho}$, showing
that $F[\bm{\rho}] \lesssim F[\bm{\rho^\prime}]$.

%
\begin{prop}
\label{prop:cg-F}
The coarse-grained Lieb internal energy functional $F: \X \to \overline{\mathbb R}$
is convex, lower semicontinuous and bounded above. The infimum in its definition
is realized.
\end{prop}

\subsection{potentials}

It is time to bring into the picture potentials and the central
relation between densities and potentials. 
We add some restrictions in a moment, but the first thing to
observe about potentials in the coarse-grained context is that
they should be uniform on cells. The point is that a 
density is a sort of atom. It contains fine-grained densities
which are equivalent under coarse-graining and we are not
permitting ourselves the resources to distinguish amongst them.
A potential that is not uniform over cells would do that, so
should be disallowed.

So, a potential will be a 4-vector function $\bm{v} = (v_0,\vec{v})$
which is uniform on cells, but we will also require them to be
respect a lower bound, in the sense that $v_0 - |\vec{v}|$ is
bounded below by a constant $-V_0$ which depends on $\Part$ and
which will be explained shortly. By contrast, potentials are
allowed to be unbounded above. The set of {\it real-valued} 4-vector
potentials obeying the lower bound is denoted $\Vc$.
We will go further, however, and allow potentials which are
{\it positive infinite}. For $\bm{v} \in \Vc$, the value of
$\bm{v}$ in a particular cell is naturally given as the real number $v_0$ 
and the vector $\vec{v}$. It can also be given as the direction
of $\vec{v}$ and the two real numbers $v_0 - |\vec{v}|$ and
$v_0 + |\vec{v}|$. 
The second method extends smoothly to
$\Vcb$, but the former is ambiguous in case $v_0 = |\vec{v}| = +\infty$.
The restriction on potentials in $\Vcb$ is thus
$-V_0 \le v_0 - |\vec{v}| \le +\infty$. 
It will be useful later in discussing topology to
recognize that the second method of specification amounts to giving
a point in $S^2 \times [-V_0,+\infty]\times[-V_0,+\infty]$, where $S^2$ 
is the two-sphere. 
Just as in the fine-grained theory, the potential energy of density 
$\bm{\rho}$ in potential $\bm{v}$ is 
$\langle {\bm v},{\bm \rho}\rangle = \int_{{\mathbb R}^3} \bm{v}\cdot\bm{\rho}$.
Decomposing ${\bm v}\cdot{\bm\rho}$ into
positive and negative parts as
$({\bm v}\cdot{\bm\rho})^\pm = \mathrm{max}(0,\pm{\bm v}\cdot{\bm\rho})$,
we see that this integral has a well-defined value in
$[-V_0 \|\rho_0\|,+\infty]$.
In the presence of external potential $\bm{v}$, the total energy of
$\bm{\rho}$ consists of internal energy and potential energy:
\beq
E[\bm{v},\bm{\rho}] \defeq F[\bm{\rho}] + \langle \bm{v},\bm{\rho}\rangle.
\nonumber
\eeq

For ${\bm \rho}\in \X^+$ and ${\bm v} \in \Vcb$, it means the same
thing to say that
${\bm \rho}$ is a {\it ground-state density}
for ${\bm v}$ or that ${\bm v}$ is a {\it representing potential}
for ${\bm \rho}$. This relation $\Rc$ between densities and potentials
is defined by
\begin{align}
{\bm \rho} \Rc {\bm v} & \Leftrightarrow \label{R-def} \\
& E[ {\bm v},{\bm\rho}] = 0, \,\, \mathrm{and}
\nonumber \\
& \forall {\bm \rho^\prime} \in 
\X^+ \,.\,
E[ {\bm v},{\bm\rho}^\prime] \ge 0.
\nonumber
\end{align}
At times it will be convenient to use the more suggestive notation
\beq
\Potl(\bm{\rho}) \defeq \{\bm{v}\in\Vcb: \bm{\rho}\,\Rc\,\bm{v}\},
\eeq
and say that $\bm{\rho}$ is {\it VREP} (``V-representable'') if
$\Potl(\bm{\rho}) \not = \emptyset$.
The basic idea of the definition is transparent:
in presence of ${\bm v}$, 
there is a state $\Gamma$
with $\pi\, \Dns\, \Gamma = {\bm\rho}$ with total
(intrinsic plus external potential) energy zero,
but no state with lower energy.
What is perhaps not immediately obvious is how this works
out correctly when improperly normalized densities are involved,
and why the ground state energy is required to be zero.
Since $v_0$ can be shifted by an overall constant 
without changing anything physical, this stipulation fixes that 
constant. It also makes everything work out for improperly normalized
densities since both the internal energy and external potential
energy scale with the overall normalization so that total energy
zero (respectively, greater than zero) is preserved under a change
of normalization.

Turn now to the origin of the lower bound $-V_0$.
Because the cells of $\Part$ are non-infinitesimal, there is a state 
of finite internal energy $N V_0$ for which all particles are in $\Omega_i$.
Thus, if ${v_0} - |\vec{v}| < V_0$ in $\Omega_i$, then that state would
have total energy {\it less than zero}. Thus, our convention on the 
constant offset of potentials actually implies the floor, 
\beq
v_0 - |\vec{v}| \ge -V_0.
\label{V-floor}
\eeq
Imposing the bound is therefore essentially not an additional restriction, 
but just a making-explicit of that fact that potentials which violate it
are not needed.

As mentioned earlier, a potential in $\Vcb$ may be viewed as a map 
from $\Part$ into ${\cal Y} = S^2 \times [-V_0,+\infty] \times [-V_0,+\infty]$.
$[-V_0,+\infty]$ is a compact space with a neighborhood basis
of $+\infty$ consisting of intervals $(a,+\infty]$. 
${\cal Y}$, and then $\Vcb \simeq {\cal Y}^\Part$, are thus also
compact as the products of compact spaces.
A basic open set for $\Vcb$ consists of the specification, for
each of a {\it finite} number of cells, of an open set in ${\cal Y}$
into which the potential must fall.
But a simpler characterization might be to say that convergence of a 
sequence of potentials amounts simply to cell-wise convergence (in ${\cal Y}$).

The next Section will show that every density in $\X^{\oplus\oplus}$ is VREP.


\section{V-representability and regularity}
\label{VREP}

This section is the core of the paper and contains
all the basic existence and regularity results.
In the next section, the uniqueness question is
taken up.

\subsection{baby steps}
\label{baby}

In \S \ref{basic-DFT}, we noted that
${\rm epi}\, F = 
\left\{ ({\bm \rho}^\prime, y) \in \X \times {\mathbb R}
: y \ge  F[{\bm \rho}^\prime]\right\}$
is convex and closed, but despite that, a straightforward
Hahn-Banach argument is not available to prove V-representability
of even one density because $\mathrm{dom}\, F$ has empty interior.
This problem is common to the conventional and coarse-grained 
theories. However, in the latter, a restriction to a fixed
finite volume allows us to make the argument work.
Control of the simplified problem will serve as a foundation
on which to build a general solution.

First, we recall the simple finite-dimensional result we need.
Let $f$ be a convex real-valued function on a convex domain 
$U \subset {\mathbb R}^n$ and $x$ be an interior point of $U$. 
Then, there exists a non-vertical plane containing the point
$(x,f(x))$ and lying on or below the graph of $f$.
Phrased algebraically, this says that there exists 
$y \in {\mathbb R}^n$ such that
$f(x) + (x^\prime - x)\cdot y \le f(x^\prime)$.

Now, suppose $\bm{\rho}\in \X^\oplus$ and $\mathrm{supp}\, \bm{\rho}$ is bounded 
(i.e., contains only a finite number of cells of $\Part$).
The set of densities with support contained in
$\mathrm{supp}\, {\bm \rho}$ is $U$, and the restriction of $F$ to $U$ is $f$.
$f$ has an upper bound given by inequality (\ref{F-bound}),
and ${\bm \rho}$ is in the interior of $U$, since it is not spin-saturated.
Thus, all the conditions are satisfied to apply the separation 
theorem of the previous paragraph.
Therefore, there is a potential $\bm{v}$ specified in $\mathrm{supp}\, \bm{\rho}$
such that (\ref{R-def}) holds, except that the test density
$\bm{\rho^\prime}$ is restricted to $U$. Since densities in $U$ are zero
on the complement of $\mathrm{supp}\, \bm{\rho}$, it does not matter 
for this what values $\bm{v}$ takes there. 
But, if we set $v_0 = +\infty$, $\vec{v} = 0$ outside 
$\mathrm{supp}\, \bm{\rho}$ then $\bm{\rho}\, \Rc\, \bm{v}$.
Thus, densities in $\X^\oplus$ with bounded support are VREP.

\subsection{hyperfinite VREP bootstrap}
\label{bootstrap}

The modest result of the previous subsection can now
be put into a nonstandard setting by means of the
Transfer principle. Then it can be used to approximate
the infinite-volume VREP problem, delivering the goods
we are after.
This technique of strengthening a modest result
by means of an excursion through the hyperfinite, which is a
common nonstandard argument style, might be called the
{\it hyperfinite bootstrap}.
Since every density in $\X^\oplus$ with bounded support is
VREP, the Transfer Principle assures us that every density
in $\Star{\X^\oplus}$ with hyperbounded support is $^\star$VREP,
and we shall see that we can adequately approximate any 
standard density by such a one. 

We begin working toward that with a finer analysis of the
potential energy $\langle \bm{v},\bm{\rho}\rangle$.
Denoting the positive/negative potential energy density maps
$({\bm v}\cdot{\bm\rho})^\pm = \mathrm{max}(0,\pm{\bm v}\cdot{\bm\rho})$
by $\PED^\pm$, the following emerges:
\begin{align}
& \X^+ \times \Vcb \xrightarrow{{\PED}^-} \ell_1^+ 
\xrightarrow{\int} {\mathbb R}^+,
\nonumber \\
& \X^+ \times \Vcb \xrightharpoondown{\PED^+} \overline{\ell^+} 
\xrightharpoondown{\int} \overline{{\mathbb R}^+}.
\label{PE-density-maps}
\end{align}
The second arrow in each sequence is just integration to
get the total negative/positive potential energy.
Both arrows in the first line are continuous and
those in the second line are lower semicontinuous
(hence the single-barbed arrows).
$\PED^-$ maps into $\ell_1^+$,
the space of $\Part$-measurable non-negative integrable 
functions topologized by means of the $L_1$ norm.
Only integrability is non-trivial, and that is a result of
the lower bound (\ref{V-floor}), as is the fact that $\PED^-$
is continuous. Continuity of $\int: \ell_1^+ \mapsto {\mathbb R}^+$
is clear.
The positive potential energy is trickier.
$\overline{\ell^+}$ denotes the space of $\Part$-measurable
functions valued in $[0,+\infty]$, equipped with the product
topology (similar to the one on $\Vcb$). 
What is intended by ``lower semicontinuity'' of
$\PED^+: \X^+\times\Vcb \rightharpoondown \overline{\ell^+}$ is
that the value in each cell of $\Part$ is lower semicontinuous.
To see that this holds, a standard argument works fine.
Since the $L^1$ topology on $\X^+$ is finer than the product
topology, it suffices to show that 
$\bm{v^\prime}(x)\cdot\bm{\rho^\prime}(x)$ is close to
or larger than $\bm{v}(x)\cdot\bm{\rho}(x)$ if
$\bm{\rho^\prime}(x)$ is close enough to $\bm{\rho}(x)$ 
and $\bm{v^\prime}(x)$ to $\bm{v}(x)$. That is 
straightforward (the ``greater than'' option is needed
for $\bm{\rho}=0$, $|\bm{v}|=\infty$).

For the other arrow, take $f \approx g \in \overline{\ell^+}$.
With $B(L)$ denoting the collection of cells with centers within distance $L$
of the origin,
$\int g\, dx = \lim_{L\to\infty} \int_{B(L)} g\, dx$ in $[0,+\infty]$.
Clearly, however, $\int_{B(L)} f \approx \int_{B(L)} g\, dx$ for limited $L$.
Thus, given $\epsilon \gg 0$, for $L$ big enough,
$\int f\, dx \ge \int_{B(L)} f\, dx \gtrsim \int_{B(L)} g \, dx 
\ge \int g\, dx - \epsilon$. Since $\epsilon$ is arbitrary,
$\int f\, dx \gtrsim \int g \, dx$. 
For handy reference, we record the conclusions in a Lemma.
\begin{lem}
\label{lem:PED-semicts}
With reference to (\ref{PE-density-maps}),
$\PED^-$ and $\int \circ \PED^-$ are continuous, 
whereas
$\PED^+$ and $\int \circ \PED^+$ are lower semicontinuous.
Hence, $\langle \Std{\bm{v}},\Std{\bm{\rho}}\rangle
 \lesssim \langle \bm{v},\bm{\rho}\rangle$ for any
$\bm{v}$, $\bm{\rho}$.
Also, if ${\bm\rho} \in \Star{\X^+}$ has bounded support, then
$\bm{v} \mapsto \langle \cdot,{\bm\rho}\rangle : \Vcb \to {\mathbb R}\cup\{+\infty\}$
is S-continuous ($\bm{v^\prime}\approx\bm{v} \Rightarrow 
\langle\bm{v^\prime},\bm{\rho}\rangle \approx
\langle\bm{v^\prime},\bm{\rho}\rangle$).
\end{lem}
\begin{proof}
Only the last statement has not been proven, but it 
reduces to the case of $\bm{\rho}$ supported in
a single cell, which is easy.
\end{proof}

From the just-proven lower semicontinuity of the potential energy 
and (Prop. \ref{prop:cg-F}) of the internal energy
for nearstandard $\bm{\rho}$,
\beq
E[ \Std{\bm v},\Std{\bm\rho}] \lesssim E[\bm{v},{\bm \rho}].
\label{ineq:E-lsc}
\eeq

\begin{prop}
\label{prop:R-closed}
$\Rc$ is closed (as a subset of $\X^+ \times \Vcb$).
\end{prop}
\begin{proof}
We show that $\mathrm{st}\, \Star{\Rc} = \Rc$. This is the
nonstandard characterization of closedness. (Note, this does
not mean that $\Star{\Rc}$ is nearstandard. Non-NS points have
no standard part.)
Thus, assume that $({\bm\rho},{\bm v}) \in \Star{\Rc}$.
Just a few lines back, in (\ref{ineq:E-lsc}), we showed that
\beq
E[ \Std{\bm v},\Std{\bm\rho}]
\lesssim E[\bm{v},{\bm \rho}] = 0.
\label{ineq:R-closed-1}
\eeq

To prove the Proposition, then, it suffices to show that
\beq
\forall {\bm\rho^\prime} \in \X^+ \, . \,\,
0 \le E[ \Std{\bm v},{\bm\rho^\prime}].
\label{R-right}
\eeq

Thus, for a contradiction, assume that there is ${\bm\rho^\prime}\in \X^+$ 
such that
$E[\Std{\bm v},{\bm\rho^\prime}] < 0$.
Since if any density satisfies the inequality, one with bounded
support (use truncation) does so, we may assume that
${\bm \rho^\prime}$ has bounded support.
Since by Lemma \ref{lem:PED-semicts},
$\langle \Std{\bm v},{\bm\rho^\prime}\rangle \approx
\langle {\bm v},{\bm\rho^\prime}\rangle$,
we obtain
$E[{\bm v},{\bm\rho^\prime}] < 0$,
and this contradicts the assumption $({\bm\rho},{\bm v}) \in \Star{\Rc}$.
\end{proof}

Note that in the course of that proof, we showed that if
${\bm\rho}$ is nearstandard and $\Star{\mathrm{VREP}}$,
then $F[\Std{\bm\rho}] \approx F[{\bm\rho}]$.
Thus, once it has been established that every
density in $\Star{\X^{+}}$ is $\Star{\mathrm{VREP}}$,
it will immediately follow that $F$ is continuous.

\begin{cor}
\label{cor:universal-VREP}
Every density in $\X^{+}$ is VREP.
\end{cor}
\begin{proof}
Let ${\bm\rho^\prime}$ be the truncation of $\Star{\bm\rho}$
at some illimited radius. Thus, ${\bm\rho^\prime}$ has 
hyperbounded support. If $\bm{\rho^\prime}$ is anywhere spin-saturated,
an additional infinitesimal perturbation will put it in $\Star{\X^\oplus}$.
By Transfer of the result in \S \ref{baby}, 
${\bm\rho^\prime}$ is $\Star{\mathrm{VREP}}$.
Thus, there is some ${\bm v}$ in $\Star{\Vcb}$ such that
${\bm\rho^\prime} (\Star{\Rc}) {\bm v}$. Therefore,
since $\Std{\bm\rho^\prime} = {\bm\rho}$,
Prop. \ref{prop:R-closed} shows that
${\bm\rho}\, {\Rc}\, (\Std{\bm v})$. 
\end{proof}

\begin{cor}
\label{F-cts}
$F$ is continuous on $\X^+$.
\end{cor}
\begin{proof}
The proof was anticipated in the remark preceding Cor. \ref{cor:universal-VREP}.
\end{proof}

\subsection{Potential Energy Density and Functional Derivatives}
\label{PED+fnctl-derivatives}

The next Lemma has additional information about the potential energy.
Part (b) will be needed for Prop. \ref{prop:PED-closed} and part (c) for
Prop. \ref{prop:fnctl-derivatives}.

\begin{lem} 
\label{lem:utilities-2}
Suppose $\bm{\rho}\, \Star{\Rc} {\bm v}$ for
nearstandard density ${\bm\rho} \in \Star{\X^+}$
and ${\bm v} \in \Star{\Vcb}$.
\hfill\break
(a) For any nearstandard $\bm{\rho^\prime} \in \Star{\X^+}$,
and illimited $L$ (superscript `$c$' denotes complement),
\beq
\int_{B(L)^c}(\bm{v}\cdot\bm{\rho^\prime})^- \, dx \approx 0.
\eeq
\hfill\break
(b) For illimited $L$,
\beq
\int_{B(L)^c}(\bm{v}\cdot\bm{\rho})^+ \, dx \approx 0.
\eeq
Hence, $\bm{v}\cdot\bm{\rho}$ is $L^1$-NS.
\hfill\break
(c)
If $\bm{\rho} + \bm{\delta\rho}\in \Star{\X^+}$ is nearstandard,
\begin{align}
^\circ\left(\int (\bm{v}\cdot\bm{\delta\rho})^- \,dx\right)  & =
\int {}^\circ(\bm{v}\cdot\bm{\delta\rho})^- \,dx =
\int ({}^\circ\bm{v}\cdot{}^\circ\bm{\delta\rho})^-\, dx,
\nonumber  \\
{}^\circ\left(\int (\bm{v}\cdot\bm{\delta\rho})^+ \,dx\right)  & 
\ge \int {}^\circ(\bm{v}\cdot\bm{\delta\rho})^+\, dx 
\ge \int ({}^\circ\bm{v}\cdot{}^\circ\bm{\delta\rho})^+\, dx.
\nonumber
\end{align}
\end{lem}
\begin{proof}
(a) If the conclusion fails, then set $\bm{\rho^{\prime\prime}}$ equal to
$\bm{\rho^\prime}$ wherever $\bm{v}\cdot{\bm{\rho^\prime}} < 0$ outside $B(L)$,
and zero everywhere else. Then, $\langle \bm{v},\bm{\rho^{\prime\prime}}\rangle \ll 0$,
while $\| \rho^{\prime\prime}_0 \| \approx 0$, so that $F[\bm{\rho^{\prime\prime}}] \approx 0$.
But that would contradict $\bm{\rho} \Rc \bm{v}$.
\hfill\break
(b) Again, if the assertion is wrong, $\bm{\rho}$ can be truncated outside
$B(L)$ to give a lower total energy in $\bm{v}$, which is a contradiction.

For nearstandardness of $\bm{v}\cdot\bm{\rho}$, take $\epsilon \gg 0$.
$\int_{B(L)^c} |\bm{v}\cdot\bm{\rho}\, dx < \epsilon$ for any illimited
$L \in \Star{\mathbb N}$. But since the inequality is an internal property
of $L$, it holds for large enough limited $L$. Thus, $\bm{v}\cdot\bm{\rho}$
is within $\epsilon$ of its truncation to $B(L)$, and the latter is certainly
$L^1$. Now take a sequence of $\epsilon_n \to 0$ and use completeness.
\hfill\break
(c)
\begin{align}
I^\pm(L) & \defeq \int_{B(L)} (\bm{v}\cdot\bm{\delta\rho})^\pm\, dx
\nonumber \\
\tilde{I}^\pm(L) & \defeq \int_{B(L)} {}^\circ(\bm{v}\cdot\bm{\delta\rho})^\pm\, dx
\nonumber
\end{align}
Then for $L \in {\mathbb N}$,
\beq
\tilde{I}^\pm(L) + \epsilon > I^\pm(L) > \tilde{I}^\pm(L) - \epsilon.
\label{Ipm-inequality}
\eeq
As in part (b), we see that this must continue to hold for small enough
illimited $L$. Now, recalling that $\Std{\bm{\rho}}\, \Rc\, \Std{\bm{v}}$,
so use parts (a) and (b) to add in integrals over $B(L)^c$ to those 
in (\ref{Ipm-inequality}).
For ``minus'', we obtain
$\int (\bm{v}\cdot\bm{\delta\rho})^-\, dx 
\approx \int {}^\circ(\bm{v}\cdot\bm{\delta\rho})^- \, dx$.
And, for ``plus'',
$\int (\bm{v}\cdot\bm{\delta\rho})^+\, dx 
\approx \int {}^\circ(\bm{v}\cdot\bm{\delta\rho})^+ \, dx$.
\end{proof}

The closedness of $\Rc$ is a property which gives some of what 
continuity would.
Suppose $\bm{\rho}_n$ is a sequence of densities in $\X^+$ converging
to $\bm{\rho}$, and $\bm{v}_n \in \Potl(\bm{\rho}_n)$. 
Since $\Vcb$ is compact, $\bm{v}_n$ has a convergent subsequence.
Because $\Rc$ is closed, this limit must be in $\Potl(\bm{\rho})$.
Put another way, for any neighborhood $U$ of $\Potl(\bm{\rho})$,
there is a neighborhood $W$ of $\bm{\rho}$ such that 
$\Potl(\bm{\rho^\prime}) \subset U$ whenever $\bm{\rho^\prime} \in W$.
This property of a set-valued map is known as
{\it upper semicontinuity}\cite{Aubin+Ekeland,Schirotzek,Aubin+Frankowska}.

The ground-state potential energy density is another set-valued map
which is closely related to $\Potl$. It is defined as
\beq
\PED(\bm{\rho}) \defeq \left\{ \bm{v}\cdot\bm{\rho}:\, \bm{v} \in \Potl(\bm{\rho}) \right\}.
\label{PED-def}
\eeq
Thus, the members of $\PED(\bm{\rho})$ are the potential energy densities
associated with $\bm{\rho}$ and the potentials which have $\bm{\rho}$ as
a ground-state density. Lemma \ref{lem:utilities-2}(b) says that $\PED$
takes values in $\ell_1$:
\beq
\X^+ \stackrel{\PED}{\rightrightarrows} \ell_1
\nonumber
\eeq

Referring to the convergent sequence 
$(\bm{\rho}_n,\bm{v}_n) \to (\bm{\rho},\bm{v})$
of the previous paragraph, Lemma \ref{lem:PED-semicts} tells us that
$\bm{v}\cdot\bm{\rho} \le \liminf \bm{v}_n\cdot\bm{\rho}_n$ pointwise.
But the fact that this sequence is in $\Rc$ does not even enter that
conclusion. Can we do better in that case? In fact, $\mathrm{Graph}\, \PED$
is also closed. This follows easily from the closedness of $\Rc$ once
we settle a tricky point with a Lemma.
\begin{lem}
\label{lem:PED-std-parts}
If $\bm{\rho}\in\Star{\X^+}$ is ns and $\bm{\rho}\, (\Star{\Rc})\, \bm{v}$,
then the $L^1$ standard part of $\bm{v}\cdot\bm{\rho}$ is given by
$\Std(\bm{v}\cdot\bm{\rho}) = \Std{\bm{v}}\cdot\Std{\bm{\rho}}$.
\end{lem}
\begin{proof}
That $\bm{v}\cdot\bm{\rho}$ is $L^1$ nearstandard was shown in
Lemma \ref{lem:utilities-2}(b). The standard part is clearly given
by taking the standard part of the value on each cell.
That $\Std({\bm{v}}\cdot{\bm{\rho}}) = \Std{\bm{v}}\cdot\Std{\bm{\rho}}$
is trivial --- except for one case. The only way it could fail is if
in some cell $\Omega$, $|\bm{v}| \approx +\infty$ and $\bm{\rho} \approx 0$,
yet $\bm{v}\cdot\bm{\rho} \gg 0$.
We will show that is not consistent with $\bm{\rho}\,{\Star{\Rc}}\bm{v}$.
Assuming that it does happen, let $\gamma$ with $\Dns \gamma = \bm{\rho}$ 
be a state realizing the infimum
defining $F[\bm{\rho}]$, take $h$ to be a function equal to $1$ outside 
$\Omega$ but going smoothly to $0$ near the center of $\Omega$, and modify
$\gamma$ by multiplying by $\prod_{i=1}^N h(x_i)^2$.
This modification costs no (Coulomb) interaction energy, at most an
infinitesimal kinetic energy from gradients of $h$, and at most an
infinitesimal potential energy increase by changing the density outside
$\Omega$. But, if $h$ is chosen appropriately, the particle number in
$\Omega$ can be reduced by (say) half, 
resulting in an appreciable (non-infinitesimal) drop of potential energy. 
But that means that $\bm{\rho}$ is not really 
a ground-state density for $\bm{v}$.
\end{proof}
\begin{prop}
\label{prop:PED-closed}
$\mathrm{Graph}\, \PED$ is closed in $\X^+\times \ell_1$.
\end{prop}
\begin{proof}
Let $\bm{\rho} \in \Star{\X^+}$ be nearstandard, and 
$\bm{v}$ be an element of $\Star{\Potl}(\bm{\rho})$.
Then $\Std{\bm{v}} \in \Potl(\Std{\bm{\rho}})$, and it
suffices to show that $\Std{\bm{v}}\cdot\Std{\bm{\rho}}$ is
the $L^1$ standard part of $\bm{v}\cdot\bm{\rho}$.
But that is precisely what Lemma \ref{lem:PED-std-parts} asserts.
\end{proof}
Referring again to our convergent sequence 
$(\bm{\rho}_n,\bm{v}_n) \to (\bm{\rho},\bm{v})$ in $\Rc$, we now see
that the potential energy density is convergent, pointwise and in total.

Although the convex analysis result that got things started in \S \ref{baby}
involved a geometrical interpretation of potentials,
that picture subsequently faded into the background. 
We now bring it forward again in the guise of 
{\it functional derivatives}. The heuristic theory of DFT suggests
that the representing potential should be equal to minus the
functional derivative of the internal energy. 
For ${\bm\rho}, {\bm\rho}+\bm{\delta\rho}\in \X^{+}$,
convexity of $F$ at least guarantees existence of the directional derivative
\beq
F^\prime[\bm{\rho};\bm{\delta\rho}] \defeq 
\lim_{\epsilon \downarrow 0}{1 \over \epsilon} \left( 
F[\bm{\rho} + \epsilon \bm{\delta\rho}] - F[\bm{\rho}] \right).
\nonumber
\eeq
By the very defintion of $\Potl({\bm\rho})$, 
\beq
\langle -{\bm v}, \bm{\delta\rho} \rangle \le F^\prime[{\bm\rho};\bm{\delta\rho}],
\nonumber
\eeq
for any ${\bm v}$ in $\Potl({\bm\rho})$. 
The result we are working toward says that if we take the maximum over
$\Potl(\bm{\rho})$ here, the inequality can be replaced by equality:
$
F^\prime[{\bm\rho};\bm{\delta\rho}] = \mathrm{max} \, 
\left\{
\langle -{\bm v}, \bm{\delta\rho} \rangle : {\bm v}\in\Potl({\bm\rho})
\right\}$.
Thus, all directional derivatives are determined in a well-defined way by
$\Potl({\bm\rho})$, and in this sense it makes sense to say that $\Potl$ is
minus the functional derivative of $F$.

\begin{prop}
\label{prop:fnctl-derivatives}
For ${\bm \rho}$ and ${\bm\rho}^\prime = {\bm\rho}+ \bm{\delta\rho}$ in 
$\X^{+}$, there is ${\bm v} \in \Potl({\bm\rho})$ satisfying
\beq
\langle -{\bm v}, \bm{\delta\rho} \rangle = F^\prime[{\bm\rho};\bm{\delta\rho}].
\nonumber
\eeq
Hence,
\beq
F^\prime[{\bm\rho};\bm{\delta\rho}] = \mathrm{max} \, 
\left\{
\langle -{\bm v}, \bm{\delta\rho} \rangle : {\bm v}\in\Potl({\bm\rho})
\right\}.
\label{functional-derivative}
\eeq
\end{prop}
\begin{proof}
Choose an infinitesimal $\epsilon > 0$.
Since (one-sided) directional derivatives of $F$ exist,
\beq
F[{\bm\rho}+\epsilon \bm{\delta\rho}] - F[\bm{\rho}]
= \epsilon ( F^\prime[\bm{\rho}; \bm{\delta\rho}] +\eta ),
\nonumber
\eeq
where $\eta \approx 0$.
Now, ${\bm\rho}+\epsilon \bm{\delta\rho}$ is $^\star$VREP by
Cor. \ref{cor:universal-VREP} and Transfer, so let 
${\bm v}\in \Star{\Vcb}$ be an element of $\Potl({\bm\rho}+\epsilon \bm{\delta\rho})$.
Then,
\beq
F[{\bm\rho}+\epsilon \bm{\delta\rho}]
\le F[\bm{\rho}]
+ \langle {\bm v}, \epsilon \bm{\delta\rho}\rangle.
\nonumber
\eeq
Combining the previous two displayed equations yields
$\langle {\bm v}, \bm{\delta\rho}\rangle 
\lesssim - F^\prime[\bm{\rho}; \bm{\delta\rho}]$.
On the other hand, by Prop. \ref{prop:R-closed}
$\Std {\bm v}\in \Potl(\bm{\rho})$, implying that
$- F^\prime[\bm{\rho}; \bm{\delta\rho}]
\le \langle \Std{\bm v}, \bm{\delta\rho}\rangle$.
Thus (remember, $\bm{\delta\rho}$ is standard),
\beq
\langle {\bm v}, \bm{\delta\rho}\rangle \approx
{}^\circ\langle {\bm v}, \bm{\delta\rho}\rangle \le
 - F^\prime[\bm{\rho}; \bm{\delta\rho}]
\le \langle \Std{\bm v}, \bm{\delta\rho}\rangle.
\label{inequality-sandwich}
\eeq
But now, since $\bm{\delta\rho}$ is standard,
Lemma \ref{lem:utilities-2}(c) says that
$^\circ\langle {\bm v}, \bm{\delta\rho}\rangle \ge
\langle \Std{\bm v}, \bm{\delta\rho}\rangle$
Thus, the inequalities in (\ref{inequality-sandwich})
collapse to equalities, showing that $\Std{\bm{v}}$ is
the potential the Corollary asserted.
\end{proof}

\section{Four-potential uniqueness and non-uniqueness}
\label{U+NU}

This section is concerned with the question of when
$\Potl({\bm \rho})$ might have more than one element.
This problem attracted renewed scrutiny in the conventional 
setting\cite{Capelle+Vignale,Eschrig+Pickett,Ullrich05,Gidopoulos07},
long after the original investigation\cite{VonBarth+Hedin}.
The conclusion reached here (Prop. \ref{NU-prop}) is similar to 
the cumulative conclusion of those works. 
But here unique continuation properties are demonstrated rather
than being assumed. In summary, in the absence of spin saturation,
the representing potential is guaranteed to (exist and) be unique 
($|\Potl (\rho)| = 1$) except in case of collinear 
magnetization in an $S_z$ eigenstate.
The next section looks more closely at the collinear case with 
spin-saturation. No nonstandard analysis is used in either of
these sections.

In contrast to previous sections, it is necessary to carefully 
consider the behavior of wavefunctions in configuration space.
Throughout this section, $\Psi$ is a fixed ground state wavefunction
of total energy $0$ in the presence of potential ${\bm v} = (u,\vec{B})$:
\beq
(H_0 + \hat{V}) \Psi = 0,
\label{SE}
\eeq
where
\beq
\hat{V} = \sum_{\alpha=1}^N \left[ u({x}_\alpha) + \vec{B}({x}_\alpha)\cdot\vec{\sigma}_\alpha \right].
\eeq
Here, $H_0$ contains kinetic energy and interaction energy and is 
diagonal in spins. The potential at the position of particle $\alpha$
is $(u(x_\alpha),\vec{B}({x}_\alpha))$. 
$\Psi$ might be a zero-energy ground state for some other potential.
Elements of $\Potl(\Psi)$ other than $(u,\vec{B})$ are generically 
denoted as $({u}^\prime,\vec{B}^\prime)$.
It will soon become clear that, in studying the non-uniqueness problem,
only the magnetic field component really requires attention.
To facilitate that, the notation
\beq
\vec{\Potl}(\bm{\rho}) = \left\{ 
\vec{B}: (u,\vec{B}) \in {\Potl}(\bm{\rho})\, \text{for some}\, u \right\}
\eeq
will be used for the projection of ${\Potl}(\bm{\rho})$.
Naturally extending previous notation, $\Potl(\Psi)$, denotes
the set of four-potentials having $\Psi$ as a zero-energy ground state. 
As previously, position space cells (``$\Part$-cells'') are denoted
$\Omega_i$, etc., $i$ being simply an abstract cell index.
Cells in configuration space (``$\Part^N$-cells'') are denoted in
the style ${\Omega}^N_{i_1\cdots i_N}$. In cell ${\Omega}^N_{i_1\cdots i_N}$,
particle $1$ is in $\Omega_{i_1}$, $\ldots$, particle $N$ is in $\Omega_{i_N}$.
The value of $(u,\vec{B})$ in cell $\Omega_i$ is denoted $(u_i,\vec{B}_i)$.
In this section and the next, instead of $(\rho_0,\vec{\rho})$ for the
4-density $\pi \Dns\, \Psi$, we use the notation $(n_i,2\vec{m}_i)$
(Note the factor of $2$).
If for every 
$(u^\prime,\vec{B}_i^\prime)\in \Potl(\Psi)$, $\vec{B}_i^\prime = \vec{B}_i$, 
then $i$ is a unique-$\vec{B}$ cell (UB cell), otherwise a
non-unique-$\vec{B}$ cell (non-UB cell). 

The following unique continuation result for single-component 
wavefunctions is fundamental.
\newtheorem*{XIII.63}{Thm. XIII.63 of RSIV\cite{RSIV}}
\begin{XIII.63}
Let $u\in H_{loc}^2$, that is, $\phi u \in D(-\Delta)$ for
each $\phi \in C_0^\infty({\mathbb R}^n)$. Let $D$ be an open
connected set in ${\mathbb R}^n$ and suppose that
\beq
|\nabla^2 u(x) | \le M |u(x)| \quad \hfill 
\label{UCP-bound}
\eeq
almost everywhere in $D$. Then, if $u$ vanishes in the neighborhood of
a single point $x_0 \in D$, $u$ is identically zero in $D$.
\end{XIII.63}

$M$ is supposed to be an arbitrary but fixed
constant. 
An examination of the proof of the basic unique continuation theorem 
shows that it extends to multicomponent wavefunctions, such as 
wavefunctions with spin indices, by replacing $| u |^2$ and 
$|\nabla^2 u|^2$ by the
corresponding sums of squares of spin components. Indeed, 
this replacement can simply be made globally throughout the 
proof of the theorem and lemmata leading up to it.
Unfortunately, that does not immediately give us any control of
individual spin components. The usual way a bound of the sort
(\ref{UCP-bound}) arises, of course, is that $u$ satisfies a Schr\"{o}dinger
equation, so that $M$ is $|V - E|$. An individual spin component might
satisfy such an equation by itself. If it is appropriately related
to a different component that does so, that can also work.

It may appear at first that the unboundedness of the Coulomb
repulsion could cause problems.
On any $\Lambda_L$, the single-particle external potential is 
bounded, so the total external potential is bounded on $\Lambda_L^N$.
The sum of the external potential and the interaction is then uniformly
bounded on $U_{L,n} = \Lambda_L^N \cap U_n$ for any $n$, where 
$U_n = \{\underline{x}: |x_i - x_j| > 1/n, \forall\, i \not=j \}$. 
Also, $U_{L,n}$ is a connected (open) set for large 
enough $n$, so the theorem will apply to it. 
But, any open subset of $\Lambda_L^N$ must have an open intersection
with $U_{L,n}$ for some $n$. So, in fact, we are free to take $D$ to be
$\Lambda_L^N$. This depends on the set where the interaction energy is
infinite having small codimension and no interior.
Our first application of the theorem is then to note that if
$\Psi\equiv 0$ on any $\Omega_{i_1\cdots i_N}^N$, then it vanishes
everywhere. Thus, there is nonzero probability to be in any
$\Part^N$-cell.

Note in passing that, by basic elliptic regularity theory\cite{Lieb+Loss},
$\Psi$ is continuous away from coincident configuration points
($x_\alpha = x_\beta$ for $\alpha \not = \beta$) and is $C^\infty$
away from coincidence or particles on the boundaries of $\Part$-cells.

If $(u^\prime,\vec{B}^\prime) \in \Potl(\Psi)$ is distinct from  $(u,\vec{B})$ 
then with
$\Delta u = u^\prime - u$ and $\Delta \vec{B} = \vec{B}^\prime - \vec{B}$,
\beq
\Delta \hat{V}\Psi = 
\sum_{\alpha=1}^N 
\left[ \Delta u({x}_\alpha) + 
\Delta\vec{B}({x}_\alpha)\cdot\vec{\sigma}_\alpha\right] \Psi = 0.
\label{Delta-v-eigenstate}
\eeq
It then follows that $[H_0 + \hat{V},\Delta\hat{V}] \Psi = 0$.
As a distributional equation on the interior of the configuration space 
cell $\Omega_{i_1 \cdots i_N}$ this yields
\beq
\sum_{\alpha=1}^N 
(\vec{B}_{i_\alpha} \times \Delta\vec{B}_{i_\alpha})\cdot
\vec{\sigma}_\alpha \Psi = 0.
\nonumber
\eeq
Now take the commutator of $H_0 + \hat{V}$ with the new
operator, to find
\beq
\sum_{\alpha=1}^N 
[\vec{B}_{i_\alpha}\times (\vec{B}_{i_\alpha} \times \Delta\vec{B}_{i_\alpha})]
\cdot \vec{\sigma}_\alpha \Psi = 0.
\nonumber
\eeq
These equations are useful in the case of $\Omega_{i i \cdots i}$,
Our basic unique continuation result assures that $\Psi$ does not vanish 
on any open set in this cell, so the equations are not vacuous. 

First, note from Eq. (\ref{Delta-v-eigenstate}) that if 
$\Delta\vec{B}_i = 0$, then $\Delta u_i = 0$ as well.
So a UB cell also has unique $u$, explaining the focus on
the UB/non-UB dichotomy.
Thus, suppose that $\Omega_i$ is a non-UB cell. 
Then, $\Delta\vec{B}_i$ can be taken nonzero, and so can
$\vec{B}_i$ (otherwise flip the roles of $\vec{B}_i$ and $\vec{B}_i^\prime$).
If $\Delta\vec{B}_i$ is not parallel to $\vec{B}_i$, that makes
$\Psi$ an eigenstate of spin along two perpendicular axes, which is
impossible. Thus, $\Delta\vec{B}_i$ is parallel to $\vec{B}_i$:
\begin{lem}
If $\Omega_i$ is a non-UB cell, then there is a unique axis
$\hat{e}_i$, such that 
for every $\vec{B}$ in $\vec{\Potl}(\Psi)$,
$\vec{B}_i$ is parallel to $\hat{e}_i$, that is,
$\vec{B}_i = {B}_i \hat{e}_i$.
\end{lem}
\begin{proof}
See previous paragraph.
\end{proof}

If $\Omega_j$ is a UB cell with $\vec{B}_j \not = 0$, there is 
also a unique axis $\hat{e}_j$, namely $\vec{B}/|\vec{B}|$.
If $\vec{B}_j = 0$ is unique, $\hat{e}_j$ is arbitrary.
Each cell $\Omega_j$ thus has its own spin quantization axis $\hat{e}_j$.
In the following, we will mostly use this {\it $\hat{e}_i$-basis}.
That is, in ${\Omega}^N_{i_1\cdots i_N}$, the component 
$\Psi_{s_1\cdots s_N} = \langle s_1\cdots s_N| \Psi \rangle$ 
has the spin of particle $\alpha$ up along 
$\hat{e}_{i_\alpha}$ if $s_\alpha = +1$ and down along 
$\hat{e}_{i_\alpha}$ if $s_\alpha = -1$.  

Consider now the restriction of $\Psi$ to
${\Omega}^N_{i_1\cdots i_N}$. The external four-potential is
diagonalized in the $\hat{e}$ basis, so the unique
continuation theorem quoted above can be applied separately to
each component $\Psi_{s_1\cdots s_N}$ to see that if
$\Psi_{s_1\cdots s_N}$ vanishes on any open set in
${\Omega}^N_{i_1\cdots i_N}$, it vanishes almost everywhere
on the cell ${\Omega}^N_{i_1\cdots i_N}$. 

At this point, we want to consider 
${\Omega}^N_{j\cdots ji_{k+1}\cdots i_N}$
with $i_{k+1},\ldots,i_N$ all different from $j$.

\begin{lem}
\label{unique-component}
If ${\Omega}_{j}$ is non-UB,
then, for fixed ${\Omega}^N_{j\cdots ji_{k+1}\cdots i_N}$
with $i_{k+1},\ldots,i_N$ all different from $j$,
the $s_{k+1}\ldots s_N$ component of $\Psi$ (in $\hat{e}$ basis)
is an eigenstate of spin along $\hat{e}_j$. 
That is, for only one $s_1,\ldots,s_k$ is $\Psi_{s_1\cdots s_N}$
nonzero.
\end{lem}

\begin{proof}
This follows straightforwardly from the eigenfunction Eq. (\ref{Delta-v-eigenstate}).
$\Delta{B}_j$ can be assumed to be nonzero along $\hat{e}_j$, giving 
\beq
k \Delta u_{j} + \Delta {B}_{j}( k - 2N_{j\downarrow})
+ \sum_{\alpha=k+1}^N \left( \Delta u_{i_\alpha} + \Delta {B}_{i_\alpha}s_\alpha
\right) = 0,
\nonumber
\eeq
where $N_{j\downarrow}$ is the number of spin-down particles in $\Omega_j$.
The equation has at most one solution for $N_{j\downarrow}$.
\end{proof}

\begin{prop}
\label{m-parallel-B}
If $\Omega_j$ is a non-UB cell, then $\hat{e}_j \times \vec{m}_j = 0$.
\end{prop}
\begin{proof}
This follows from Lemma \ref{unique-component} by summation over
cell occupations and integration.
\end{proof}

At this point we turn attention from a single configuration space cell
to a pair of neighboring cells ${\Omega}^N_{k i_2\cdots i_N}$
and ${\Omega}^N_{j i_2\cdots i_N}$.
``Neighboring'' means that $\Omega_{k}$ shares a face with
$\Omega_{j}$. The indices $i_2 \cdots i_N$ are fairly inert, so
the cell index notation is abbreviated to $\Omega^N_{k [i]}$
and $\Omega^N_{j [i]}$.

For the following argument we use a slight modification of the 
$\hat{e}$ basis, namely using $\hat{e}_{k}$ in both 
$\Omega_{k}$ and $\Omega_{j}$.
 ($\hat{e}_{k [i]}$-basis).
\begin{lem}
\label{no-flip}
Suppose that $\Omega_j$ and $\Omega_{k}$ are neighboring
cells.  Using $\hat{e}_{k[i]}$-basis
in both
${\Omega}^N_{j [i]}$ and ${\Omega}^N_{k[i]}$,
if $\Psi_{s_1\cdots s_N}$ vanishes in ${\Omega}^N_{j [i]}$,
Then it also vanishes on ${\Omega}^N_{k[i]}$.
\end{lem}
\begin{proof}
The unique continuation theorem is applicable because the potential 
is diagonalized in ${\Omega}^N_{k [i]}$, while
the condition (\ref{UCP-bound}) for our chosen component is satisfied 
in ${\Omega}^N_{j [i]}$ by hypothesis.
\end{proof}
A possibly useful mnemonic is to say that ``moving one particle 
from one cell to another cannot flip the spins of other particles'',
bearing in mind that ``moving'' does not refer to a dynamical process
but just a shift of attention.

\begin{lem}
\label{parallel-neighbors}
If $\Omega_{j}$ and $\Omega_{k}$ are neighboring $\Part$-cells
and both non-UB, then $\hat{e}_j$ and $\hat{e}_{k}$ are parallel.
\end{lem}
\begin{proof}
In the $\hat{e}$ spin basis, find an occupied 
$s[t]$ spin state in ${\Omega}^N_{j [i]}$, with all 
of $i_2,\ldots,i_N$ different from $j$ and $k$. 
Then $(-s)[t]$ is unoccupied by Lemma \ref{unique-component}.
Also, there is some
occupied $s^\prime[t]$ in ${\Omega}^N_{k [{i}]}$,
in ${\hat{e}}_{k [{i}]}$ basis.
for if there were no such $s^\prime$, then Lemma \ref{no-flip}
would guarantee that $s[t]$ is unoccupied in 
${\Omega}^N_{j [{i}]}$.
Furthermore, $(-s^\prime)[t]$ is then unoccupied in 
${\Omega}^N_{k [{i}]}$, again by Lemma \ref{unique-component}.

Thus, in ${\Omega}^N_{k [{i}]}$,
\beqa
\Psi_{(-s) [t]} &=&
\langle -s | s^\prime \rangle \Psi_{s^\prime [t]}
+ \langle -s | -s^\prime \rangle \Psi_{(-s^\prime) [t]}
\nonumber \\
&=&\langle -s | s^\prime \rangle \Psi_{s^\prime [t]}.
\nonumber
\eeqa
$\Psi_{(-s) [t]}= 0$ in ${\Omega}^N_{j [{i}]}$, and therefore
satisfies a bound of type (\ref{UCP-bound}) there.
But, it also satisfies such a bound in ${\Omega}^N_{k [{i}]}$.
This follows because $\Psi_{s^\prime [t]}$ satisfies such a bound
there by virtue of obeying a Schr\"odinger equation,
and $\Psi_{(-s) [t]}$ has a fixed proportionality to 
$\Psi_{s^\prime [t]}$ in ${\Omega}^N_{k [{i}]}$ according
to the previous display.

The basic unique continuation theorem then says that
$\Psi_{(-s) [t]}=0$ in ${\Omega}^N_{k [{i}]}$.
But that implies that $|s\rangle = |s^\prime \rangle$,
and since these spin states are eigenstates of
$\hat{e}_j\cdot\vec{\sigma}$ and $\hat{e}_{k}\cdot\vec{\sigma}$ 
respectively, $\hat{e}_j$ and $\hat{e}_{k}$ are coaxial.
\end{proof}

The picture according to what has been proven so far is of
connected components of non-UB cells, separated by UB cells.
On each connected cluster, there is a unique common $\hat{e}$ axis. 
So, it is now time to consider UB cells. It turns out that
the existence of even one has drastic consequences.

\begin{lem}
\label{spin-saturation}
If there is a UB cell, then all non-UB cells are spin-saturated.
\end{lem}

\begin{proof}
Label one of the unique-$\vec{B}$ cells `$0$'.
Then, considering the $\Part^N$-cell ${\Omega}^N_{0\cdots 0}$ with all particles in 
$\Omega_0$, the eigenfunction Eq. (\ref{Delta-v-eigenstate}) becomes $N \Delta u_0 = 0$.
So, $\Delta u_0 = 0$ also. 

Let $\Omega_j$ be a non-UB cell and look in ${\Omega}^N_{j0\cdots 0}$.
(If there are no non-UB cells, the Lemma is vacuously true.)
The argument now is like that in Lemma \ref{unique-component}:
\beq
\Delta u_j + \Delta\vec{B}_j\cdot\vec{\sigma}_1 = 0
\nonumber 
\eeq 
for any spin component 
which occurs in the wavefunction in ${\Omega}^N_{j 0\cdots0}$, where
$\Delta\vec{B}_j$ can be taken non-zero along $\hat{e}_j$.
This equation can be satisfied for only one value of 
$s^j =  \hat{e}_j\cdot\vec{\sigma}$, either $+1$ or $-1$ but not both.
Furthermore, $\Delta u_j = -s^j |\Delta \vec{B}_j|$.  
Shifting attention to 
${\Omega}^N_{jj0\cdots 0}$, ..., ${\Omega}^N_{jj\cdots j}$
in turn, the same argument shows that all the spins in $\Omega_j$ are
in the $s^j$ spin state. 

But, this argument does not tell us anything about the wavefunction in
$\Omega_{jk0 \cdots 0}$, where $\Omega_j$ and $\Omega_k$ are 
two different non-UB cells.
If $|\Delta\vec{B}_j| = |\Delta\vec{B}_k|$, it looks as though it
might be possible to flip a spin in each cell while continuing to respect 
Eq. (\ref{Delta-v-eigenstate}).
 
Lemma \ref{no-flip} comes to the rescue here. 
For example, ${\Omega}^N_{jj\cdots j}$ is saturated, by the preceding argument,
and ${\Omega}^N_{j\cdots j k}$ 
can be reached by moving (shift of attention) particle $N$ from one cell to 
another until $k$ is reached.
According to Lemma \ref{no-flip}, at no step in this process do we see
flips of the particles left in $\Omega_j$, so that the only possibility 
allowed by the eigenfunction Eq. (\ref{Delta-v-eigenstate}) is that particle $N$ ends up in state
$s^k$. It is clear how the example generalizes.

Thus, if there is even one unique-$\vec{B}$ cell, 
then any particle which is ever in the 
non-UB cell $\Omega_j$ is in spin state $s^j$. 
That shows spin saturation in cell $j$, namely, $|\vec{m}_j| = n_j/2$.
\end{proof}

From the foregoing results, the main conclusion of this section can
now be assembled. 
If there are UB (unique-$\vec{B}$) cells as well as
non-UB cells, then the latter are spin-saturated.
We reconsider spin saturated cases in the next section, but for now
assume no spin saturation, so this case is ruled out.
All cells are thus non-UB, and all $\hat{e}_j$ are
therefore along the same axis by Lemma \ref{parallel-neighbors}.
In that case, there is a {\em global} spin quantization axis
which diagonalizes the Hamiltonian. 
$S_z$, the total spin along the common axis, is a good quantum number.
Taking $\Psi$ to be an eigenstate of $S_z$, and looking at
$\Omega_{ii\cdots i}$, we see that 
$\Delta u_i = -({2 S_z/N}) \Delta B_i$
Thus, different $S_z$ eigenstates cannot share more than one potential.
Also, using this, the eigenfunction Eq. (\ref{Delta-v-eigenstate}) becomes
\beq
\sum_\alpha \Delta B_{i_\alpha} \left( -\frac{2 S_z}{N} + s_\alpha \right) = 0,
\nonumber
\eeq
which implies that all $\Delta B_i$ are equal.
Therefore, 
\begin{prop}
\label{NU-prop}
if $\bm{\rho}$ is nowhere spin-saturated but 
$\Potl(\bm{\rho})$ is not a singleton, then
$\vec{m}$ is everywhere along a common axis,
$\vec{m} = m\hat{e}$, and the total spin $\int m \, d{x} = S_z$ is a
half-integer.  Furthermore $\Potl(\bm{\rho})$ is a one-parameter family
with $\Delta\vec{B}$ only allowed to be uniform:
\beq
u_i = u_i^0 - 2S_z\Delta B/N, \quad
\vec{B}_i = \hat{e}[B_i^0 + \Delta B].
\nonumber
\eeq
\end{prop}
\begin{proof}
Preceding discussion.
\end{proof}

Note that this last result implies that
$\Potl({\bm\rho})$ can be locally determined from directional derivatives 
for variation of the density in only a single cell since such
variations can be used to determine the ranges of $u_i$ and $\vec{B}_i$
and the proposition shows how to put them together to construct all
of $\Potl(\bm{\rho})$.

\section{collinear states and spin-saturation}
\label{collinear}

This short section takes up densities which are somewhere spin-saturated, 
especially the important case of collinear spin density. 
As for the general case, it is easy to see that a spin-saturated cell is 
non-UB, so according to the remark following Lemma \ref{parallel-neighbors},
the saturated cells form clusters all magnetized in the same direction
separated by non-spin-saturated, non-UB cells.
Prop. \ref{prop:R-closed} (closedness of the graph of $\Potl$)
offers a possible route to find representing
potentials for such densities. 

The rest of this section concentrates on the collinear case.
Thus $\vec{m}$ is everywhere along the $\hat{z}$
axis, though it might be zero. Cases of spin saturation were
excluded from the general results of the previous two Sections, but 
collinear spin-saturated states clearly exist in nature, so
it is important to make special provision for them.

\begin{prop}
\label{m-along-z-implies-B-along-z}
If $\vec{m}$ is everywhere along the $z$ axis, then so is $\vec{B}$ for
any $\vec{B} \in \vec{\Potl}(\bm{\rho})$.
\end{prop}

\begin{proof}
Assume $\pi\, \Dns \Psi = (n,2\vec{m})$ with $\vec{m}$ everywhere along $\hat{z}$,
and $(u,\vec{B}) \in \Potl(n,2\vec{m})$. 
Simultaneous rotation of the spins and of $\vec{B}$ by the same angle 
about the $z$ axis preserves those relationships, as well as
$\E(\Psi)$. 
Since $\vec{m}$ is invariant under the rotation, if $\vec{B}_i$ is not,
then $\Omega_i$ is a non-UB cell. But in that case, $\vec{B}$ is
along $\hat{z}$ according to Prop. \ref{m-parallel-B}.
\end{proof}

\begin{prop}
If $\bm{\rho} = (n,2\vec{m}) \in \X^{++}$ and $\vec{m}$ is 
everywhere along the $z$ axis, then 
$\Potl (\bm{\rho}) \not = \emptyset$ if and only if $\bm{\rho}$ is either
everywhere spin-saturated or nowhere spin-saturated.
\end{prop}
\begin{proof}

The case of nowhere spin-saturated $\bm{\rho}$ has already been dealt with,
so suppose that $\bm{\rho}$ is everywhere spin-saturated with 
$\vec{m} = -\hat{z} n/2$.
 
In the spinless version of the theory of Sec. \ref{VREP},
$n = \rho_0$ is V-representable; it is the density of some ground state 
$\gamma$ in potential $v$.
Now, consider the family of four-potentials with 
$\vec{B} = B \hat{z}$, and $u - B = v$. 
The product of $\gamma$ and $| S_z = -N/2 \rangle$
is an eigenstate of any of these $\bm{v}$. Since $S_z$ is a good 
quantum number, all that needs to be done is to add a large enough
uniform constant to $B$ (and subtract it from $u$) so that the ground
states in all other $S_z$ sectors have greater energy than does 
$\gamma \otimes |S_z = - N/2 \rangle$.

To see that other cases are not V-representable, assume 
$\vec{m} = m \hat{z}$ with ${m} = - n/2$ somewhere, but not everywhere,
so that $(u,\vec{B}) \in \Potl(\bm{\rho})$ satisfies
$\vec{B} = B \hat{z}$ according to Prop. \ref{m-along-z-implies-B-along-z}.
$S_z$ is a good quantum number, so consider the ground state energies in each
$S_z$ sector. If $S_z = -N/2$ is not tied for minimum energy, the
assumed situation is impossible. If a degeneracy does occur, then the
$S_z = -N/2$ components can be removed from the state and the result is
still a ground state for $(u,\vec{B})$. But this state has zero density
in at least one cell, and that is impossible because it violates our 
basic unique continuation principle. 
\end{proof}

\section{Concluding remarks}
\label{conclusion}

This paper extends previously established\cite{Lammert06,Lammert10a,Lammert10b}
results of coarse-grained DFT to SDFT. Here is a quick summary.
As a function of coarse-grained four-density with $L^1$ norm,
the Lieb internal energy functional $F$ is continuous.
For a coarse-grained density $\bm{\rho}$, $\Potl(\bm{\rho})$ is 
the set of representing four-potentials which are constant on cells
with additive constant fixed by $F[\bm{\rho}] + \langle \bm{v},\bm{\rho}\rangle = 0$.
Each four-potential $\bm{v}$ in $\Potl(\bm{\rho})$ has a ground state
$\gamma$ with a four-density that coarse-grains to $\bm{\rho}$:
$\pi\, \Dns\, \gamma = \bm{\rho}$. If $\bm{\rho}$ is everywhere
nonzero and nowhere spin-saturated ($\bm{\rho} \in \X^{\oplus\oplus}$), 
then $\bm{\rho}$ is VREP: $\Potl(\bm{\rho}) \not= \emptyset$.
In fact, $\Potl$ is the functional derivative of $F$ in this sense: 
for $\bm{\rho}$ and $\bm{\rho} + \bm{\delta\rho}$ in $\X^{\oplus\oplus}$,
the directional derivative
$F^\prime[{\bm\rho};\bm{\delta\rho}]$
is equal to 
$\mathrm{max} \, 
\left\{
\langle -{\bm v}, \bm{\delta\rho} \rangle : {\bm v}\in\Potl({\bm\rho})
\right\}$.
Furthermore, $\bm{\rho} \mapsto \{ \bm{\rho}\cdot\bm{v}:
\bm{v} \in \Potl(\bm{\rho}) \}$ is an $L^1$ upper semicontinuous
set-valued function, $\Potl$ takes values in closed sets and
$\mathrm{Graph}\, \Potl$ is closed in $\X \times \Vc$.
Apart from spin-saturated densities, 
only for densities $\bm{\rho}$ with collinear magnetization that 
integrates to a half-integer can $\Potl(\bm{\rho})$ have more than 
one element. 
And in that case, there is a global axis $\hat{e}$ such that
$\vec{B}$ is everywhere along $\vec{e}$ for all 
$\vec{B} \in \vec{\Potl}(\bm{\rho})$
and different elements of $\vec{\Potl}(\bm{\rho})$ differ by
a uniform shift along $\vec{e}$.
Collinear states with fully saturated spin are also V-representable.

This paper has an additional
aim of promoting the use of nonstandard analysis in mathematical
and theoretical physics.
The results of \S \ref{VREP} were obtained with the aid of
nonstandard analysis tools, resulting in proofs which
are intuitive and of modest technical sophistication.
The turn to infinitesimal methods was spurred by the  
hope that it would shed light on the limit of coarse-graining
scale going to zero. In the meantime, it has shown its value
for the infinite-volume limit. 

\begin{acknowledgments}
I thank Carsten Ullrich for a stimulating conversation at the
2009 APS March Meeting and Vin Crespi for suggestions on the manuscript.
\end{acknowledgments}

%

 \end{document}